\documentstyle[12pt,epsfig]{article}
\def\shiftdown#1{#1\llap{\lower.04ex\hbox{#1}}}

\begin{document}

\begin{center}

{\bf \large
Current conservation, screening and the magnetic moment of the $\Delta$ 
resonance.}
{\footnotemark}

\vspace{2mm}

{\bf 1. Formulation without quark degrees of freedom }

\footnotetext{ 
Supported by the "Deutsche Forschungsgemeinschaft" under contract
GRK683
}

\end{center}


\vspace{1mm}

\noindent{
{\large \bf A.\ I.\ Machavariani$^a$ $^b$ $^c$ and Amand Faessler $^a$ } 

}

\vspace{1mm}

\noindent{\small

{  \rm $^a$ 
Institute\ f\"ur\ Theoretische\ Physik\ der\ Univesit\"at\
 T\"ubingen,\newline T\"ubingen\ D-72076, \ Germany}\\

{\rm $^b$ Joint\ Institute\ for\ Nuclear\ Research,\ Dubna,\ Moscow\
region\ 141980,\ Russia}\\

{\rm $^c$ High Energy Physics Institute of Tbilisi State 
University,
University str.  9,  Tbilisi 380086, Georgia  }\\




}

\vspace{0.05cm}


\medskip
\begin{abstract}
{\bf

The pion-nucleon bremsstrahlung  $\pi+N\Longrightarrow\gamma'+\pi'+N'$
is  studied in a new  form of current conservation. According
to this condition, the internal and external particle 
radiation parts of the   
$\pi N$ radiation amplitude  have opposite signs, 
i.e., they contain  terms which must cancel each other.
 Therefore, one has a screening of the internal and external particle 
radiation  in the $\pi N$ bremsstrahlung.
In particular, it is shown that  the double $\Delta$ exchange 
diagram with the $\Delta-\gamma' \Delta'$ vertex cancel against the 
appropriate longitudinal part of the external particle radiation diagrams.
Consequently, a model independent relation between
the magnetic dipole moments  of the 
$\Delta^+$ and $\Delta^{++}$ resonances and the anomalous magnetic moment
of the proton $\mu_p$ is obtained,
where $\mu_{\Delta}$ is expressed by $\mu_p$  
as $\mu_{\Delta^+}={ {M_{\Delta}}\over {m_p}} \mu_p$ 
and $\mu_{\Delta^{++}}={3\over 2}\mu_{\Delta^+}$
in agreement with the values extracted from the fit for the 
experimental cross section of the $\pi^+ p\to\gamma'\pi^+ p$ 
reaction. }

\end{abstract}


\newpage

\begin{center}
                  {\bf 1. INTRODUCTION}
\end{center}
\medskip

The $\pi N$ bremsstrahlung was extensively investigated in the past 
in order to study the electromagnetic properties  
of the $\Delta$ resonances and their form factors. 
The main reason for the determination of the 
electromagnetic moments of the $\Delta$ resonances is that on one hand, 
the $\Delta$'s  are described as a $\pi N$ resonances 
with the corresponding poles of the
$\pi N$ amplitude and, on the other hand 
 $\Delta$'s  are often treated as 
independent particles in the models of strong interaction.
In addition, the quark content of the proton is the same as for the $\Delta^+$,
and differs from the quark content of  the $\Delta^{++}$. Therefore,  
determination of the electromagnetic moments of the $\Delta$ resonances 
is important for the definition of the electromagnetic structure of the
nucleons and the $\Delta$ resonances.

In contrast to nucleons, the direct experimental measurement of the 
electromagnetic moments of the $\Delta$'s is today impossible. 
Therefore, the present experimental electromagnetic moments of the $\Delta$'s
are obtained using a fit to the experimental cross sections of the $\pi N$ 
 bremsstrahlung \cite{Leung}-\cite{Boss}.
The analysis of these data by different theoretical models yields
different magnetic moments of the $\Delta$'s.  For instance,
the magnetic dipole moment of the $\Delta^{++}$ $\mu_{\Delta^{++}}$
obtained within the framework of the low energy photon theorem 
\cite{Low}-\cite{Ding} is  $\mu_{\Delta^{++}}=4.7$-$6.9\mu_B$ \cite{Nefkens} in
 nuclear magnetons $\mu_B=e/2m_N$,  while the potential models yield
$\mu_{\Delta^{++}}=5.6$-$7.5\mu_B$ \cite{Wittman} or   
$\mu_{\Delta^{++}}=4.5\pm 0.95\mu_B$ \cite{Boss}. 
The theoretical results for different models
fitted to the experimental data 
\cite{Low}-\cite{Franklin} are
shown in Table 1 in the conclusions
of this paper. These results indicate  substantial  discrepancies 
between the different predicted values for 
$\mu_{\Delta^{++}}$ and $\mu_{\Delta^{+}}$.

In this paper an analytic and  model-independent relation 
for the magnetic moments of the $\Delta$ resonances is suggested.
This relation is
based on a new form of current conservation 
for the total on mass shell and on energy shell amplitude  
of the $\pi N$ bremsstrahlung
${\cal A}^{\mu}_{\gamma'\pi'N'-\pi N}$. The 
corresponding  current conservation
$$k'_{\mu}{\cal A}^{\mu}_{\gamma'\pi'N'-\pi N}=
k'_{\mu}{\cal E}^{\mu}_{\gamma'\pi'N'-\pi N}+
{\cal B}_{\pi'N'-\pi N}
=0\eqno(1.1a)$$
consists of the external  particle radiation   amplitude
${\cal E}^{\mu}_{\gamma'\pi'N'-\pi N}$ depicted in Fig. 1
and the sum of the off shell $\pi N\to\pi' N'$  scattering amplitudes 
 ${\cal B}_{\pi'N'-\pi N}$. 
This condition is obtained in the 
same approach as the Ward-Takahashi identities in the
usual quantum field theory \cite{BD2,IZ}.
Using current conservation for the total $\pi N$ radiation amplitude
$$k'_{\mu}{\cal A}^{\mu}_{\gamma'\pi'N'-\pi N}=
k'_{\mu}{\cal E}^{\mu}_{\gamma'\pi'N'-\pi N}+
k'_{\mu}{\cal I}^{\mu}_{\gamma'\pi'N'-\pi N}
=0.\eqno(1.2)$$
for the on shell external 
and internal particle radiation amplitudes 
 ${\cal E}^{\mu}_{\gamma'\pi'N'-\pi N}$ (Fig.1) and 
 ${\cal I}^{\mu}_{\gamma'\pi'N'-\pi N}$ (Fig.2A)
one can represent (1.1a) as

$$k'_{\mu}{\cal E}^{\mu}_{\gamma'\pi'N'-\pi N}=-k'_{\mu}{\cal
  I}^{\mu}_{\gamma'\pi'N'-\pi N}
=-{\cal B}_{\pi'N'-\pi N}\eqno(1.1b)$$ 
which determines an additional relation between the on shell external
and internal particle radiation amplitudes.

\vspace{5mm}

\begin{figure}[htb]
\includegraphics[width=14.0cm]{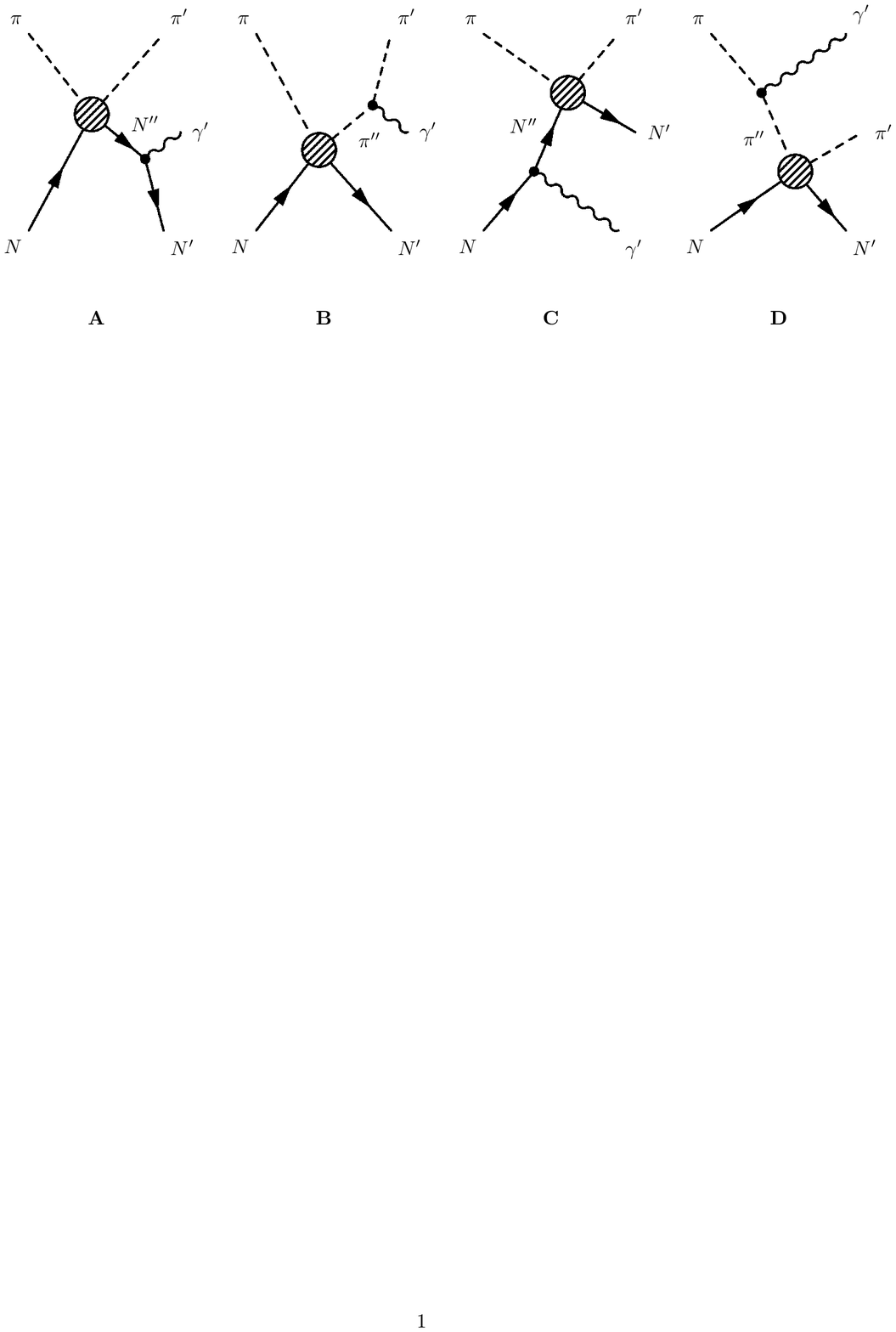}
\vspace{-15.0cm} 
\caption{{\protect\footnotesize {\it
Diagrams describing the external particle radiation amplitude 
${\cal E}_{\gamma'\pi'N'-\pi N}^{\mu}$ in (1.1a,b).
Diagrams A and C correspond to the radiation of the external 
nucleons. Diagrams B and D describe the emission of 
the photon by the external pions. 
The hatched circle indicates the off shell 
$\pi N$ elastic scattering amplitudes (2.9a,b,c,d).
$N"$ and $\pi"$ denote the intermediate nucleon and pion states.
}} }
\label{fig:one}
\end{figure}

\vspace{5mm}

Thus the problem of the validity of 
current conservation (1.1a) is reduced to the determination of the 
internal particle radiation amplitudes (Fig. 2A)  
${\cal I}^{\mu}_{\gamma'\pi'N'-\pi N}$  which satisfy 
the condition (1.1b). 

We shall show that the $\Delta$ radiation amplitude 
${\cal I}^{\mu}_{\gamma'\pi'N'-\pi N}(\Delta-\gamma\Delta)$ 
in Fig. 2B and the corresponding part of the external particle 
radiation amplitude ${\cal E}^{\mu}_{\gamma'\pi'N'-\pi N}$ (Fig. 1)
denoted as
$({\cal E_L}^{3/2})^{\mu}_{\gamma'\pi'N'-\pi N}(\Delta-\gamma\Delta)$
satisfy  current conservation

$$k'_{\mu}\biggl[{\cal A}^{\mu}_{\gamma'\pi'N'-\pi N}\biggr]
^{Projection\ on\ spin\ 3/2\ particle\  states}
_{2\Delta\ exchange\ with\ \Delta-\gamma'\Delta'\ vertex}=$$
$$k'_{\mu}
({\cal E_L}^{3/2})^{\mu}_{\gamma'\pi'N'-\pi N}(\Delta-\gamma\Delta)
+{\cal B}^{3/2}_{\pi' N'-\pi N}(\Delta-\gamma\Delta)
\eqno(1.3a)$$
or
$$k'_{\mu}({{\cal E_L}^{3/2}})^{\mu}_{\gamma'\pi'N'-\pi N}(\Delta-\gamma\Delta)
=-{k'}_{\mu}{\cal I}_{\gamma'\pi' N'-\pi N}^{\mu}(\Delta-\gamma\Delta)=
-{\cal B}^{3/2}_{\pi' N'-\pi N}(\Delta-\gamma\Delta),\eqno(1.3b)$$
where the lower index $ _{\cal L}$ and the upper index $ ^{3/2}$ denote 
the longitudinal and the spin-isospin  $(3/2,3/2)$ part of the corresponding 
amplitudes. $(\Delta-\gamma\Delta)$ indicates a $\Delta$ radiation
vertex with on mass shell $\Delta$'s in
$({{\cal E_L}^{3/2}})^{\mu}_{\gamma'\pi'N'-\pi N}$ and in 
  ${\cal B}^{3/2}_{\pi'N'-\pi N}$.

The intermediate $\Delta$'s in (1.3b)  
and in the $\Delta-\gamma\Delta$ vertex are on mass shell,
i.e. the four momentum of the $\Delta$ $P_{\Delta}$ is determined as
$P^o_{\Delta}=\sqrt{{\sf m}_{\Delta}^2+{\bf P}_{\Delta}^2} $,
where ${\sf m}_{\Delta}$ denotes the effective complex mass of the $\Delta$
which is determined by the $\Delta$ pole position of the $\pi N$ amplitude. 
In the present approach  the 
intermediate $\Delta$ radiation amplitude
${\cal I}_{\gamma'\pi' N'-\pi N}^{\mu}(\Delta-\gamma\Delta)$ (Fig. 2B)
is constructed unambiguously
using only the on mass shell $\Delta$-pole part of the $\pi N$ amplitude.   
The corresponding  3D time-ordered field theoretical construction of the 
$\Delta$ radiation amplitude
was presented  by \cite{NP,Ann,MF}. 
This approach  is generalized in appendix C for any
$s$ depending mass ${\sf m}_{\Delta}(s)$.
In particular, the  formulation considered does not use 
the effective Lagrangian with the Heisenberg operators of the 
$\Delta$. Therefore, the off mass shell $\Delta$
ambiguities does not appear.
For the 3D time-ordered representations of the diagrams in
Fig. 1 and Fig. 2B with the on mass shell intermediate pions, nucleons
and $\Delta$'s we shall use 
the following  analytic decompositions of the amplitudes 
${\cal E}^{\mu}_{\gamma'\pi'N'-\pi N}$ and ${\cal B}_{\pi'N'-\pi N}$
in (1.1a,b) in order to separate current conservation (1.3a,b):

{\bf I.} Decomposition over the nucleon and antinucleon exchange parts.

{\bf II.} Separation of the  
longitudinal and transverse parts of
${\cal E}^{\mu}_{\gamma'\pi'N'-\pi N}$ in  (1.1a).

{\bf III.} Partial wave decomposition of the off shell $\pi N$ 
amplitudes in  ${\cal E}^{\mu}_{\gamma'\pi'N'-\pi N}$
and in ${\cal B}_{\pi'N'-\pi N}$. This procedure is necessary for 
separation of the  $\Delta$ resonance 
$(3/2,3/2)$ spin-isospin states in (1.1a,b).
It also include  projections on the intermediate spin $3/2$ states in 
the $\gamma N-N$ and  $\gamma \pi-\pi$ vertices.

{\bf IV.} Separation of the current conservation conditions with and without
$\Delta$-pole terms in the off mass shell $\pi N$ amplitudes.

{\bf V.} Reproduction of the double $\Delta$ exchange amplitudes
using a sum of the $\Delta$-pole terms in 
${\cal E}^{\mu}_{\gamma'\pi'N'-\pi N}$ and in 
${\cal B}_{\pi'N'-\pi N}$. In the final 
${\cal E}^{\mu}_{\gamma'\pi'N'-\pi N}(\Delta-\gamma\Delta)$ 
the $\Delta$ radiation  vertex has the same form as 
the  usual $\Delta-\gamma\Delta$ vertex.

\vspace{5mm}

\begin{figure}[htb]
\includegraphics[width=10.0cm]{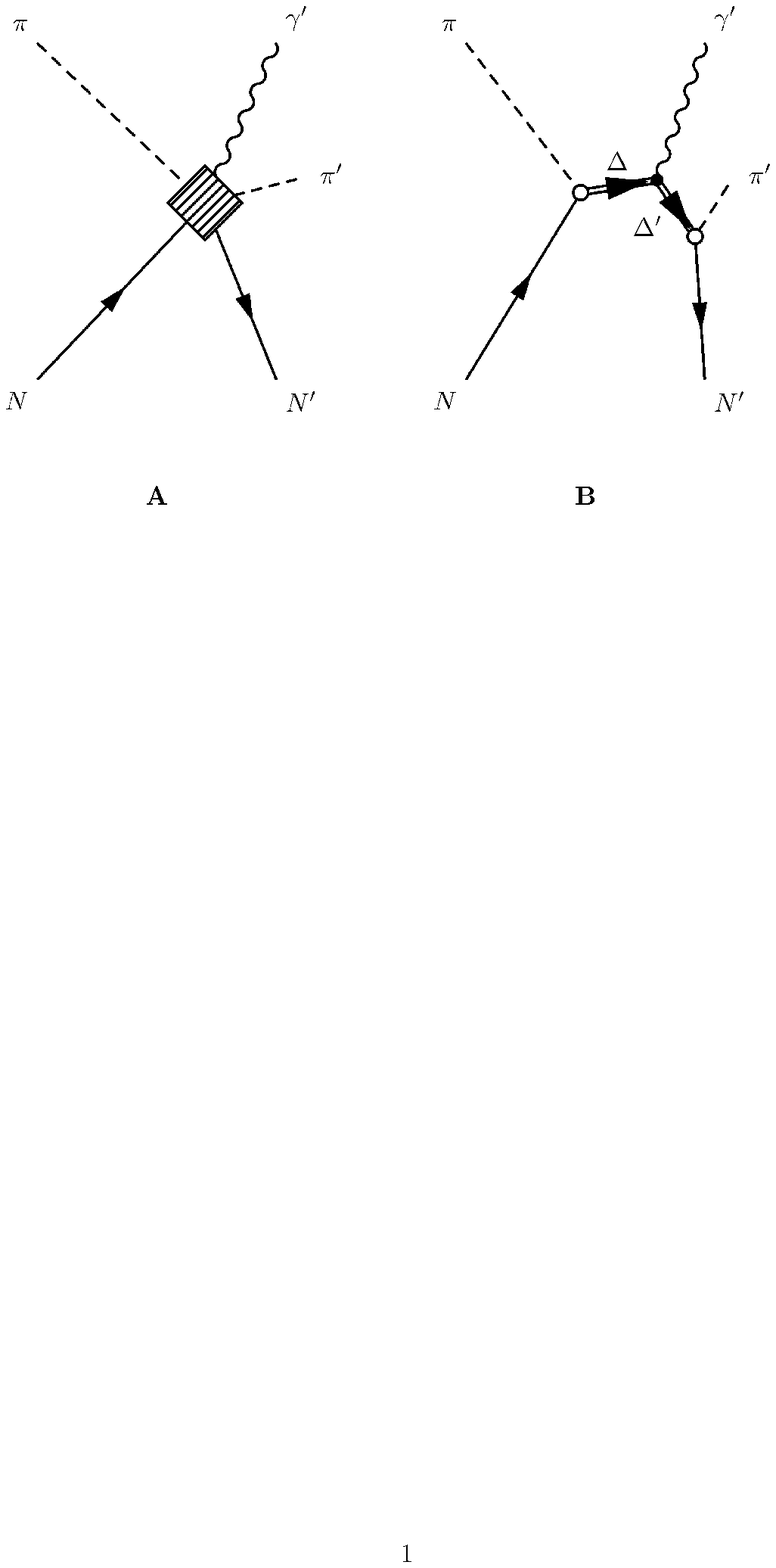}
\vspace{-14.0cm} 
\caption{{\protect\footnotesize {\it
 Diagram A presents  a symbolic description of the internal particle  
radiation amplitude ${\cal I}_{\gamma'\pi'N'-\pi N}^{\mu}$
in (1.1a,b). A special part of the amplitude 
${\cal I}_{\gamma'\pi'N'-\pi N}^{\mu}$ in diagram A is given 
by diagram  B which describes the double $\Delta$ exchange
amplitude ${\cal I}^{\mu}_{\gamma'\pi'N'-\pi N}(\Delta-\gamma\Delta)$
with the photon emission from the intermediate $\Delta$.
The $\Delta-\gamma\Delta$ vertex with on mass shell $\Delta$'s (3.7)
in ${\cal I}^{\mu}_{\gamma'\pi'N'-\pi N}(\Delta-\gamma\Delta)$
contains the magnetic dipole moment $\mu_{\Delta}$ of the $\Delta$ 
(see appendix B).   The unambiguous construction of 
${\cal I}^{\mu}_{\gamma'\pi'N'-\pi N}(\Delta-\gamma\Delta)$ within
the 3D time-ordered field theoretical approach is given in 
appendix C.
}}}
\label{fig:two}
\end{figure}

\vspace{5mm}

An important property of 
${\cal I}^{\mu}_{\gamma'\pi'N'-\pi N}(\Delta-\gamma\Delta)$
is that it satisfies not 
the separate current conservation condition 
$${k'}_{\mu}{\cal I}_{\gamma'\pi' N'-\pi N}^{\mu}(\Delta-\gamma\Delta)
\ne 0,\eqno(1.4)$$
because the $\Delta-\gamma'\Delta'$ vertex consists of  the  intermediate 
 $\Delta$ four moments $P_{\Delta}^{\mu}$, ${P'}_{\Delta}^{\mu}$
and ${k'}_{\Delta}^{\mu}=P_{\Delta}^{\mu}-{P'}_{\Delta}^{\mu}\ne k'_{\mu}$ 
(see appendix B).

Current conservation (1.3a,b) can be reinforced
if  one takes into account that only 
$({\cal E_L}^{3/2})^{\mu}_{\gamma'\pi'N'-\pi N}(\Delta-\gamma\Delta)$
and ${\cal I}^{\mu}_{\gamma'\pi'N'-\pi N}(\Delta-\gamma\Delta)$ have the 
same double $\Delta$ exchange poles and the same analytical structure of the 
$\Delta-\gamma\Delta$ vertex. Therefore,

$$({\cal E_L}^{3/2})^{\mu}_{\gamma'\pi'N'-\pi N}(\Delta-\gamma\Delta)=-
{\cal I}^{\mu}_{\gamma'\pi'N'-\pi N}(\Delta-\gamma\Delta).\eqno(1.5)$$

This relation allows to determine the magnetic dipole
 moment of the $\Delta$ resonances.
Thus, from the equality of the vertex functions in    
${\cal I}^{\mu}_{\gamma'\pi'N'-\pi N}(\Delta-\gamma\Delta)$
and ${\cal E}^{\mu}_{\gamma'\pi'N'-\pi N}(\Delta-\gamma\Delta)$
which contain  $\mu_{\Delta^+}$
and  $\mu_p$ correspondingly, it follows that $\mu_{\Delta^+}$ is analytically 
defined by $\mu_p$.

The important property of  (1.5) is the equality and  
cancellation of the intermediate $\Delta$ radiation term  
${\cal I}^{\mu}_{\gamma'\pi'N'-\pi N}(\Delta-\gamma\Delta)$ 
(Fig. 2B) and the corresponding longitudinal
part of the external particle radiation amplitudes
$({\cal E_L}^{3/2})^{\mu}_{\gamma'\pi'N'-\pi N}(\Delta-\gamma\Delta)$.
This equality and cancellation are a part of 
the general screening of the internal particle
terms via the sum of the external particle radiation diagrams 
in Fig. 1 because other parts 
of ${\cal E}^{\mu}_{\gamma'\pi'N'-\pi N}$ and 
${\cal I}^{\mu}_{\gamma'\pi'N'-\pi N}$ in  (1.1b)
are also equal and cancel each other.

Current conservation  (1.1a,b) has the same form as 
in the approach  based on the Low theorem (or low energy photon theorem)
for the reactions with
 soft photons \cite{Mink}-\cite{Lin}. 
Unlike the present formulation,
these approaches  start from the external particle 
radiation amplitude ${\cal E}_{\gamma'\pi'N'-\pi N}^{\mu}$ (Fig. 1)
which determines the full $\pi N$
bremsstrahlung amplitude
in the infrared energy region of the emitted photon $(k'\to 0)$.
One can reproduce the double $\Delta$ exchange amplitude
using the sum of the $\Delta$-pole $\pi N$ amplitudes in 
${\cal E}^{\mu}_{\gamma'\pi'N'-\pi N}$
(Fig. 1) as it was noted in \cite{Mink}{\footnotemark}  and was applied  in  
numerous other papers (see \cite{Ding})
within the low energy photon approach \cite{Low}-\cite{Lin}.
This approach is based on a approximation
 ${\cal A}^{\mu}_{\gamma'\pi'N'-\pi N}\Longrightarrow
{\cal E}^{\mu}_{\gamma'\pi'N'-\pi N}+
{\cal I}^{\mu}_{\gamma'\pi'N'-\pi N}(\Delta-\gamma\Delta)$ which allows
to calculate the cross sections of the $\pi N$ bremsstrahlung  
for the extraction of the magnetic dipole
moment of the $\Delta$. But in this approach
the equality and cancellation of 
${\cal I}^{\mu}_{\gamma'\pi'N'-\pi N}(\Delta-\gamma\Delta)$ and
${\cal E}^{\mu}_{\gamma'\pi'N'-\pi N}(\Delta-\gamma\Delta)$ according to
(1.5) was not taken into account.
Moreover,  the recipe 
of construction of the bremsstrahlung amplitude in the low energy photon 
limit $k'\to 0$ is not unique due to the 
 ambiguities of the low energy photon approximations \cite{Ding}.

\footnotetext{
The relationship between the
external particle 
radiation diagrams in Fig. 1 and the double $\Delta$ exchange term in Fig. 2B
were used  in \cite{Ding} based on the Brodsky-Brown 
 identities \cite{Brodsky1,Brodsky2} for the diagrams
in the tree approximation.}

Unlike the low energy photon approach \cite{Low}-\cite{Lin},
the present approach is not restricted  to
 the infrared energy region of the emitted photon $(k'\to 0)$,
i.e.,  (1.1a,b) and (1.5) are exactly valid for any energy of the final
photon. Moreover, in
the present approach 
the electromagnetic form factors of the $\Delta$'s 
are determined through the $\Delta$-pole 
residues of the off shell $\pi N$ amplitudes.  
The suggested formulation
can be applied to other reactions with  
conserved current like pion photo-production reaction, 
Compton scattering, processes
with external vector $\rho$ or $\omega$ mesons etc.

This paper consists of four sections and three appendices. 
Current conservation  (1.1a,b)   
for the on shell bremsstrahlung amplitudes are derived in  Section 2.
In this section  the equations (1.1a,b) are decomposed into  
independent current conservations with one nucleon and 
one antinucleon intermediate states.
The following chain of the decompositions
${\cal E}^{\mu}_{\gamma'\pi'N'-\pi N}\Longrightarrow...\Longrightarrow
({\cal E_L}^{3/2})^{\mu}_{\gamma'\pi'N'-\pi N}(\Delta-\gamma\Delta)$, 
 ${\cal B}_{\pi'N'-\pi N}
\Longrightarrow...\Longrightarrow 
{\cal B}^{3/2}_{\pi'N'-\pi N}(\Delta-\gamma\Delta)$ 
with the final form of current conservation (1.3a,b)
is given  in Section 3 and in Appendix A. 
The derivation of  (1.5) with the extraction of the
magnetic dipole moment of the $\Delta^+$ and $\Delta^{++}$ resonances
is given in Sec. 3.
The conclusions and the comparison with the magnetic dipole moments
of other authors (Table 1)  are presented in Sec. 4.
In Appendix B 
construction of the 
$\Delta-\gamma'\Delta'$ vertices
with the on
mass shell $\Delta$'s is considered. 
Reproduction of the double $\Delta$ exchange diagram in Fig. 2B
within the usual  time-ordered 
field-theoretical approach for the $\pi N$ bremsstrahlung amplitude is
given in Appendix C.



\vspace{0.25cm}

\begin{center}
                  {\bf 2. Ward-Takahashi identities for the 
pion-nucleon bremsstrahlung amplitude  }
\end{center}

\vspace{0.25cm}

\par
We consider the radiative pion-nucleon scattering 

$$\pi(p_{\pi})\ +\ N(p_{N})\Longrightarrow \gamma'(k')\ +\ \pi'(p'_{\pi})
\ +\ N'({p'_N})$$
with the on mass shell momentum 
of the  pions ($p_{\pi}=(\sqrt{ {\bf p}_{\pi}^2+m_{\pi}^2},{\bf p}_{\pi})$,
${p'}_{\pi}=(\sqrt{ {\bf p'}_{\pi}^2+m_{\pi}^2},{\bf p'}_{\pi})$,  nucleons
($p_{N}=(\sqrt{ {\bf p}_{N}^2+m_{N}^2},{\bf p}_{N})$,
${p'}_{N}=(\sqrt{ {\bf p'}_{N}^2+m_{N}^2},{\bf p'}_{N})$, and final photon
(${k'}^2=0$).
The energy-momentum of the emitted photon
is $k'_{\mu}=(p_N+p_{\pi}-p'_{\pi}-p'_N)_{\mu}$.

Following  the  derivation of the Ward-Takahashi 
identities (see e.g. ch. ${\bf 8.4.1}$ in
 the book of Itzykson and Zuber\cite{IZ}) we start with 
the on shell amplitude 
$A_{\gamma'\pi' N'-\pi N}^{\mu}$

$${k'}_{\mu}A_{\gamma'\pi' N'-\pi N}^{\mu}
({\bf p'_{\pi},p'_N,k';p_{\pi},p_N})=$$
$${\overline u}({\bf p'}_N)
(\gamma_{\nu} {p'_N}^{\nu}-m_N)({p'_{\pi}}^2-m_{\pi}^2)
{k'}_{\mu}{\tau}^{\mu}
(\gamma_{\nu} {p_N}^{\nu}-m_N)({p_{\pi}}^2-m_{\pi}^2)
 u({\bf p}_N),\eqno(2.1)$$
where the Green function ${\tau}^{\mu}$ is expressed 
via the photon source operator ${\cal J}^{\mu}(z)$ and 
the pion and  nucleon 
field operators $\Phi(x)$ and $\Psi(y)$ as

$$
{k'}_{\mu}{\tau}^{\mu}=
i\int d^4zd^4y'd^4x'd^4yd^4xe^{ik'z+ip'_{\pi}x'+ip'_Ny'-ip_{\pi}x-ip_Ny}$$
$${{\partial}\over{\partial z^{\mu}} }
<0|{\sf T}\biggl(\Psi(y')\Phi(x'){\cal J}^{\mu}(z)
{\overline \Psi}(y)\Phi^+(x)\biggr)|0>.\eqno(2.2a)$$

In this paper we use the same definition and normalization
for the Dirac spinors as in  \cite{IZ}.
In particular, $u({\bf p}_N)$ denotes the spinor of the nucleon with
the three-momentum ${\bf p}_N$.

We shall use the well known relation for the time-ordered product 
of the quantum field operators

$${{\partial}\over{\partial z^{\mu}} }
<0|{\sf T}\biggl(\Psi(y')\Phi(x'){\cal J}^{\mu}(z)
{\overline \Psi}(y)\Phi^+(x)\biggr)|0>=
<0|{\sf T}\biggl(\Psi(y')\Phi(x')
\Bigl({{\partial}\over{\partial z^{\mu}} }{\cal J}^{\mu}(z)\Bigr)
{\overline \Psi}(y)\Phi^+(x)\biggr)|0>$$
$$+\delta(z_o-x'_o)<0|{\sf T}\biggl(\Psi(y')
\biggl[{\cal J}^{o}(z),\Phi(x')\biggr]
{\overline \Psi}(y)\Phi^+(x)\biggr)|0>$$
$$+\delta(z_o-y'_o)<0|{\sf T}\biggl(\Phi(x')
\biggl[{\cal J}^{o}(z),\Psi(y')\biggr]
{\overline \Psi}(y)\Phi^+(x)\biggr)|0>$$.
$$+\delta(z_o-x_o)<0|{\sf T}\biggl(\Psi(y')\Phi(x')
\biggl[{\cal J}^{o}(z),\Phi^+(x)\biggr]
{\overline \Psi}(y)\biggr)|0>$$
$$+\delta(z_o-y_o)<0|{\sf T}\biggl((\Psi(y')\Phi(x')
\biggl[{\cal J}^{o}(z),{\overline \Psi}(y)\biggr]
\Phi^+(x)\biggr)|0>\eqno(2.2b)$$

and the equal-time commutation conditions  

$$\biggl[ {\cal J}^{o}(z),\Psi(y')
\biggr]\delta(z_o-y'_o)=-e_{N'}\delta^{(4)}(z-y')\Psi(y');\ \ \
\biggl[ {\cal J}^{o}(z),{\overline \Psi}(y)
\biggr]\delta(z_o-y_o)=e_N\delta^{(4)}(z-y){\overline \Psi}(y)\eqno(2.3a)$$

$$\biggl[ {\cal J}^{o}(z),\Phi(x')
\biggr]\delta(z_o-x'_o)=-e_{\pi'}\delta^{(4)}(z-x')\Phi(x');\ \ \
\biggl[ {\cal J}^{o}(z),{ \Phi^+}(x)
\biggr]\delta(z_o-x_o)=e_{\pi}\delta^{(4)}(z-x){ \Phi^+}(x),\eqno(2.3b)$$
where $e_{N'}$, $e_{\pi'}$, $e_N$ and  $e_{\pi}$ stand for the charge  
of the nucleons and pions in the final and initial states. In particular,
$e_N=e,0$ for the proton and the neutron, and $e_{\pi}=\pm e,0$ for pions.

After substitution of  (2.3a,b) in 
 (2.2a) and  integration over $d^4z$ 
we obtain

$${k'}_{\mu}{\tau}^{\mu}=
-i\int d^4y'd^4x'd^4yd^4xe^{ip'_{\pi}x'+ip'_Ny'-ip_{\pi}x-ip_Ny}
\biggl(e_{N'}e^{ik'y'}+e_{\pi'}e^{ik'x'}-
e_{N}e^{ik'y}-e_{\pi}e^{ik'x} \biggr)$$
$$<0|T\biggl(\Psi(y')\Phi(x')
{\overline \Psi}(y)\Phi^+(x)\biggr)|0>.\eqno(2.2c)$$

Equal-time commutators (2.3a,b) are the result of the commutation relations
 between the electric charge operator $Q=\int d^3x {\cal J}^{o}(x)$ and the 
particle field operators with the charge $e$. 
These conditions 
express electric charge conservation for the local fields,
i.e., they represent one of the first principles in quantum field theory. 

Substituting (2.2c) into (2.1) we get

$${k'}_{\mu}A_{\gamma'\pi' N'-\pi N}^{\mu}
({\bf p'_{\pi},p'_N,k';p_{\pi},p_N})=-i
(2\pi)^4\ \delta^{(4)}(p'_N+p'_{\pi}+k'-p_{\pi}-p_N)$$
$$\Biggl[{\overline u}({\bf p'}_N)(\gamma_{\nu} {p'_N}^{\nu}-m_N)
{ {e_{N'}}\over{\gamma_{\nu} (p'_N+k')^{\nu}-m_N  }}
<out;{\bf p'}_{\pi}|J(0)|{\bf p}_{\pi}{\bf p}_N;in>$$
$$+( {p'}_{\pi}^2-m_{\pi}^2){{e_{\pi'}}\over{(p'_{\pi}+k')^2-m_{\pi}^2} }
<out;{\bf p'}_{N}|j_{\pi'}(0)|{\bf p}_{\pi}{\bf p}_N;in>$$
$$-
<out;{\bf p'}_{\pi}{\bf p'}_{N}|{\overline J}(0)|{\bf p}_{\pi};in>
{{e_{N}}\over{\gamma_{\nu} (p_N-k')^{\nu}-m_N} }(\gamma_{\nu} {p_N}^{\nu}-m_N)
u({\bf p}_N)$$
$$-<out;{\bf p'}_{\pi}{\bf p'}_{N}|j_{\pi}(0)|{\bf p}_{N};in>
{{e_{\pi}}\over{(p_{\pi}-k')^2-m_{\pi}^2}}
(p_{\pi}^2-m_{\pi}^2)
\Biggr]\eqno(2.4)$$
where $J(x)=(i\gamma_{\nu}\partial/\partial x_{\nu}-m_N)\Psi(x)$
and 
$j_{\pi}(x)=(\partial^2/\partial x^{\nu}\partial x_{\nu}+m_{\pi}^2)\Phi(x)$
denote the source operator of the nucleon and the pion.

For the on mass shell external particles 
${k'}_{\mu}A_{\gamma'\pi' N'-\pi N}^{\mu}$
(2.4) vanishes. In particular, for $k'=0$\ \ \  
${k'}_{\mu}A_{\gamma'\pi' N'-\pi N}^{\mu}$  disappears 
due to cancellation of the on shell $\pi N$ amplitudes
in (2.4). Thus expression (2.4) presents
current conservation 
for the on shell bremsstrahlung amplitude

$${k'}_{\mu}
\Biggl[ A_{\gamma'\pi' N'-\pi N}^{\mu}
({\bf p'_{\pi},p'_N,k';p_{\pi},p_N})
\Biggr]_{on\ mass\ shell\ \pi',\ N',\ \pi,\ N}=0.\eqno(2.5)$$

It is convenient to extract the full energy-momentum conservation 
$\delta$ function from the  radiative $\pi N$ scattering 
amplitude $A_{\gamma'\pi' N'-\pi N}^{\mu}$ and introduce the corresponding
amplitude
$<out;{\bf p'}_{N}{\bf p'}_{\pi}|{\cal J}^{\mu}(0)|{\bf p}_{\pi}{\bf p}_N;in>$

$${k'}_{\mu}A_{\gamma'\pi' N'-\pi N}^{\mu}=-i
(2\pi)^4\ \delta^{(4)}(p'_N+p'_{\pi}+k'-p_{\pi}-p_N)
{k'}_{\mu}
<out;{\bf p'}_{N}{\bf p'}_{\pi}|{\cal J}^{\mu}(0)|{\bf p}_{\pi}{\bf p}_N;in>,
\eqno(2.6)$$

Afterwards, using the identity $a/(a+b)\equiv 1- b/(a+b)$
in  (2.4) 
$\biggl($ i.e. 
$(\gamma_{\nu} {p'_N}^{\nu}-m_N)
/\Bigl({\gamma_{\nu} (p'_N+k')^{\nu}-m_N  }\Bigr)=1-\gamma_{\mu} k'^{\mu}/
\Bigl({\gamma_{\nu} (p'_N+k')^{\nu}-m_N  }\Bigr)$;\ \ \ 
$ ({p'}_{\pi}^2-m_{\pi}^2)/\Bigl( (p'_{\pi}+k')^2-m_{\pi}^2\Bigr)=1-
2{k'}^{\mu}(p'+k')_{\mu}/\Bigl( (p'_{\pi}+k')^2-m_{\pi}^2\Bigr)$ etc.$\biggr)$
 we obtain

$$
{k'}_{\mu}
<out;{\bf p'}_{N}{\bf p}_{\pi'}|{\cal J}^{\mu}(0)|{\bf p}_{\pi}{\bf p}_N;in>=
{\cal B}_{\pi' N'-\pi N}+{k'}_{\mu}{\cal E}_{\gamma'\pi' N'-\pi N}^{\mu}
=0,\eqno(2.7)$$

where $p_{N}+p_{\pi}-p_{N'}-p_{\pi'}-k'=0$ and

$${\cal B}_{\pi' N'-\pi N}=
e_{N'}{\overline u}({\bf p'}_N)
<out;{\bf p'}_{\pi}|J(0)|{\bf p}_{\pi}{\bf p}_N;in>
+e_{\pi'}<out;{\bf p'}_{N}|j_{\pi'}(0)|{\bf p}_{\pi}{\bf p}_N;in>$$
$$-e_N<out;{\bf p'}_{\pi}{\bf p'}_{N}|{\overline J}(0)|{\bf p}_{\pi};in>
u({\bf p}_N)-e_{\pi}<out;{\bf p'}_{\pi}{\bf p'}_{N}|j_{\pi}(0)|{\bf p}_{N};in>,
\eqno(2.8a)$$

$${\cal E}_{\gamma'\pi' N'-\pi N}^{\mu}=
-\Biggl[{\overline u}({\bf p'}_N)\gamma^{\mu}
{{\gamma_{\nu} (p'_N+k')^{\nu}+m_N }\over{
2p'_Nk'}}e_{N'}
<out;{\bf p'}_{\pi}|J(0)|{\bf p}_{\pi}{\bf p}_N;in>$$
$$+(2{p'}_{\pi}+k')^{\mu}{{e_{\pi'}}
\over{ {2p'_{\pi}k' }} }
<out;{\bf p'}_{N}|j_{\pi'}(0)|{\bf p}_{\pi}{\bf p}_N;in>$$
$$-e_N
<out;{\bf p'}_{\pi}{\bf p'}_{N}|{\overline J}(0)|{\bf p}_{\pi};in>
{ {\gamma_{\nu} (p_N-k')^{\nu}+m_N}
\over{2p_Nk'} }\gamma^{\mu}
u({\bf p}_N)$$
$$-<out;{\bf p'}_{\pi}{\bf p'}_{N}|j_{\pi}(0)|{\bf p}_{N};in>
{{e_{\pi}}\over{2p_{\pi}k'} }
(2p_{\pi}-k')^{\mu}\Biggr]\eqno(2.8b)$$

The  identity (2.7) is derived for the on shell  
total $\pi N$ bremsstrahlung amplitude (2.6).
This identity   consists of ${\cal E}_{\gamma'\pi' N'-\pi N}^{\mu}$
(2.8b), which has the
form of the external particle radiation diagrams
 in Fig. 1, and ${\cal B}_{\pi' N'-\pi N}$ (2.6a), which consists of
 the sum of the different off mass shell $\pi N$ amplitudes.
 (2.7) is derived using
the same technique as for the well-known
Ward-Takahashi identity \cite{IZ}. But the usual 
Ward-Takahashi identity connects the off mass shell
$n+1$ and $n$ point vertices and Green functions. 
For the on mass shell external
particles the usual Ward-Takahashi  identity transforms 
into (2.4), which generates (2.7) 
using the simple algebraic identity $a/(a+b)\equiv 1- b/(a+b)$.
 Therefore, we designate (2.7) as the modified
Ward-Takahashi identity for the on shell $\pi N$ radiation
amplitude 
$<out;{\bf p'}_{N}{\bf p}_{\pi'}|{\cal J}^{\mu}(0)
|{\bf p}_{\pi}{\bf p}_N;in>$.

It must be noted  that  (2.4) and (2.8b) 
do not contain the full electromagnetic 
form factors of the external particles as follows
from the equal-time commutation rules (2.3a,b). 
In particular, 
${\cal E}_{\gamma'\pi' N'-\pi N}^{\mu}$ (2.8b) consists of the 
incomplete $\gamma NN$ and $\gamma \pi\pi$ vertex functions  $e_N\gamma^{\mu}$
and $e_{\pi}({p'}_{\pi}^{\mu}+p_{\pi}^{\mu})$
with the physical charges and off shell $\pi N$ amplitudes.  
The amplitude (2.8b)
can be described via a sum of the corresponding Feynman diagrams in Fig.1.
${\cal E}_{\gamma'\pi' N'-\pi N}^{\mu}$ with the full electromagnetic 
vertices was the basic  expression in
 derivation of the  low energy photon  theorem \cite{Low}.
Various applications of this method are given in 
\cite{Adler,Mink,Ding,Liou2,Lin}.
The external particle radiation amplitudes in Fig. 1 are responsible for the
infrared  behavior of the total bremsstrahlung amplitude,
i.e., in the low energy photon limit they represent the leading
diagrams. 
The present derivation of current conservation (2.7) 
based on the general condition (2.2c), i.e., (2.7)
is not restricted by the limit $k'=|{\bf k'}_{\gamma}|\to 0$.

The off shell $\pi N$ amplitudes in (2.8a,b) 
are functions of three on mass shell moments 
from which one can construct only three  independent 
Lorentz-invariant (Mandelstam) variables.
Therefore, we have

$${\overline u}({\bf p'}_N)
<out;{\bf p'}_{\pi}|J(0)|{\bf p}_{\pi}{\bf p}_N;in>
={\cal T}_{N'}\biggl( (p'_{\pi}-p_{\pi}-p_N)^2;s,t_{\pi}\biggr)=
{\cal T}_{N'}\biggl( m_N^2+2k'p'_N;s,t_{\pi}\biggr)
\eqno(2.9a)$$

$$<out;{\bf p'}_{N}|j_{\pi'}(0)|{\bf p}_{\pi}{\bf p}_N;in>
={\cal T}_{\pi'}\biggl( (p'_{N}-p_{\pi}-p_N)^2;s,t_{N}\biggr)=
{\cal T}_{\pi'}\biggl( m_{\pi}^2+2k'p'_{\pi};s,t_{N}\biggr)\eqno(2.9b)$$

$$<out;{\bf p'}_{\pi}{\bf p'}_{N}|{\overline J}(0)|{\bf p}_{\pi};in>
u({\bf p}_N)
={\cal T}_{N}\biggl(s',t_{\pi};(p'_{\pi}+p'_N-p_{\pi})^2\biggr)=
{\cal T}_{N}\biggl(s',t_{\pi}; m_N^2-2k'p_N\biggr)\eqno(2.9c)$$

$$<out;{\bf p'}_{\pi}{\bf p'}_{N}|j_{\pi}(0)|{\bf p}_{N};in>
={\cal T}_{\pi}\biggl(s',t_N;(p'_{\pi}+p'_N-p_{\pi})^2\biggr)=
{\cal T}_{\pi}\biggl(s',t_N; m_{\pi}^2-2k'p_{\pi}\biggr).\eqno(2.9d)$$

The four moments of the fourth off mass shell particle
in the $\pi N$ amplitudes (2.9a,b,c,d) are determined via 
the energy-momentum  
conservation for the bremsstrahlung amplitude
$<out;{\bf p'}_{N}{\bf p'}_{\pi}|{\cal J}^{\mu}(0)|{\bf p}_{\pi}{\bf p}_N;in>$
(2.6) with $p'_N+p'_{\pi}+k'=p_{\pi}+p_N$,
i.e., the off shell behavior of these $\pi N$ amplitudes is defined by 
$k'_{\gamma}$. The related invariant variables are

$$s'=(p'_N+p'_{\pi})^2;\ \ \ s=(p_N+p_{\pi})^2=(p'_N+p'_{\pi}+k')^2
=s'+2k'(p'_N+p'_{\pi})=s'+2k'(p_N+p_{\pi})\eqno(2.10a)$$ 

$$t_N=(p'_N-p_N)^2;\ \ \ t_{\pi}=(p'_{\pi}-p_{\pi})^2\eqno(2.10b)$$

with the following relations between them: 

$$t_{\pi}+(p'_{\pi}-p_N)^2+s
=m_{\pi}^2+2m_{N}^2+(p'_{\pi}-p_{\pi}-p_N)^2,\eqno(2.11a)$$

$$t_N+(p'_N-p_{\pi})^2+s=m_N^2+2m_{\pi}^2+(p'_N-p_{\pi}-p_N)^2,\eqno(2.11b)$$

$$t_{\pi}+(p_{\pi}-p'_N)^2+s'
=m_{\pi}^2+2m_{N}^2+(p_{\pi}-p'_{\pi}-p'_N)^2,\eqno(2.11c)$$

$$t_N+(p_N-p'_{\pi})^2+s'=m_N^2+2m_{\pi}^2+(p_N-p'_{\pi}-p'_N)^2.
\eqno(2.11d)$$

Next we rewrite expressions (2.8a,b) in the time-ordered three-dimensional
form, where the particle and antiparticle contributions in the
intermediate states are separated. Using 
the completeness conditions
$u({\bf p'_N\pm k'}){\overline u}({\bf p'_N\pm k'})
+ v({\bf p'_N\pm k'}){\overline v}({\bf p'_N\pm k'})={\sf 1}$
for the intermediate one nucleon state, we obtain  

$${\cal B}_{\pi' N'-\pi N}(N)
=e_{N'}{\overline u}({\bf p'}_N)
u({\bf p_N'+k'}){\overline u}({\bf p_N'+k'})
<out;{\bf p'}_{\pi}|J(0)|{\bf p}_{\pi}{\bf p}_N;in>$$
$$+e_{\pi'}<out;{\bf p'}_{N}|j_{\pi'}(0)|{\bf p}_{\pi}{\bf p}_N;in>
-e_N<out;{\bf p'}_{\pi}{\bf p'}_{N}|{\overline J}(0)|{\bf p}_{\pi};in>
u({\bf p_N-k'}){\overline u}({\bf p_N-k'})
u({\bf p}_N)$$
$$-e_{\pi}<out;{\bf p'}_{\pi}{\bf p'}_{N}|j_{\pi}(0)|{\bf p}_{N};in>,
\eqno(2.12a)$$

$${\cal E}_{\gamma'\pi' N'-\pi N}^{\mu}(N)=
-\Biggl[(2{p'}_{\pi}+k')^{\mu}{ {e_{\pi'}}
\over{ {2p'_{\pi}k' }} }
<out;{\bf p'}_{N}|j_{\pi'}(0)|{\bf p}_{\pi}{\bf p}_N;in>
$$
$$+{{
{\overline u}({\bf p'}_N)\Bigl[(2p'_N+k')^{\mu}
-i\sigma^{\mu\nu}k'_{\nu}\Bigr] }\over{ 2p'_Nk'} }
u({\bf p_N'+k'}){\overline u}({\bf p_N'+k'})
 e_{N'}<out;{\bf p'}_{\pi}|J(0)|{\bf p}_{\pi}{\bf p}_N;in>
$$
$$
-e_N<out;{\bf p'}_{\pi}{\bf p'}_{N}|{\overline J}(0)|{\bf p}_{\pi};in>
u({\bf p_N-k'}){\overline u}({\bf p_N-k'})
{ {\Bigl[(2p_N-k')^{\mu}-i\sigma^{\mu\nu}k'_{\nu}
\Bigr]u({\bf p}_N) }\over{2p_Nk'} }$$
$$-<out;{\bf p'}_{\pi}{\bf p'}_{N}|j_{\pi}(0)|{\bf p}_{N};in>
{{e_{\pi}}\over{2p_{\pi}k'} }
(2p_{\pi}-k')^{\mu} 
\Biggr],\eqno(2.12b)$$

where $(N)$ indicates the part of 
${\cal E}_{\gamma'\pi' N'-\pi N}^{\mu}$ with the  
one-nucleon propagator. The corresponding part of 
${\cal B}_{\pi' N'-\pi N}$ is denoted  as
${\cal B}_{\pi' N'-\pi N}(N)$.  
Similarly, for the intermediate antinucleon part of (2.8a,b) we have

$${\cal B}_{\pi' N'-\pi N}({\overline N})
=-e_{N'}{\overline u}({\bf p'}_N)
v({\bf p'_N+k'}){\overline v}({\bf p'_N+k'})$$
$$+e_N<out;{\bf p'}_{\pi}{\bf p'}_{N}|{\overline J}(0)|{\bf p}_{\pi};in>
v({\bf p_N-k'}){\overline v}({\bf p_N-k'})
u({\bf p}_N),\eqno(2.13a)$$

$${\cal E}_{\gamma'\pi' N'-\pi N}^{\mu}({\overline N})=
 {{ {\overline u}({\bf p'}_N)\Bigl[(2p'_N+k')^{\mu}
-i\sigma^{\mu\nu}k'_{\nu}\Bigr] }\over{ 2p'_Nk'} }
v({\bf p'_N+k'}){\overline v}({\bf p'_N+k'})
 e_{N'}<out;{\bf p'}_{\pi}|J(0)|{\bf p}_{\pi}{\bf p}_N;in>
$$
$$
-e_N<out;{\bf p'}_{\pi}{\bf p'}_{N}|{\overline J}(0)|{\bf p}_{\pi};in>
 v({\bf p_N-k'}){\overline v}({\bf p_N-k'})
{ {\Bigl[(2p_N-k')^{\mu}-i\sigma^{\mu\nu}k'_{\nu}
\Bigr]u({\bf p}_N) }\over{2p_Nk'} }.\eqno(2.13b).$$

For the derivation of  (2.12a,b) and (2.13a,b)  the simple
relations of the Dirac spinors 
${\overline u}({\bf p'}_N)\gamma^{\mu} 
( \gamma_{\nu} {p'_N}^{\nu}+m_N)=
{\overline u}({\bf p'}_N)2{ p'_N}^{\mu}$,
 ${\overline u}({\bf p'}_N)\gamma^{\mu} 
\gamma^{\nu} k'_{\nu}=
{k'}^{\mu}{\overline u}({\bf p'}_N)-i{\overline u}({\bf p'}_N)
\sigma^{\mu\nu}k'_{\nu}$ were used.

Expressions (2.13a,b) contain the intermediate 
$\pi\to\pi'N'{\overline  N}$ transition amplitude. 
Using the identity 
$k'_{\mu}\Bigl[(P+P')^{\mu}-i\sigma^{\mu\nu}k'_{\nu}\Bigr]=s-s'$
it is easy to obtain that

$$k'_{\mu}{\cal E}_{\gamma'\pi' N'-\pi N}^{\mu}({\overline N})
+{\cal B}_{\pi' N'-\pi N}({\overline N})=0.\eqno(2.14a)$$

Consequently, instead of the full Ward-Takahashi identity  (2.7) we get

$${k'}_{\mu}
<out;{\bf p'}_{N}{\bf p}_{\pi'}|{\cal J}^{\mu}(0)|{\bf p}_{\pi}{\bf p}_N;in>=
{\cal B}_{\pi' N'-\pi N}(N)+
{k'}_{\mu}  {\cal E}_{\gamma'\pi' N'-\pi N}^{\mu}(N)
=0.\eqno(2.14b)$$

The gauge terms proportional to $k'_{\mu}$  in (2.12b)
do not contribute to the $\pi N$ bremsstrahlung
amplitude because  the product of polarization vector
$\epsilon^{\mu}({\bf k'},\lambda)$ of the final photon and $k'_{\mu}$
vanishes $\epsilon^{\mu}({\bf k'},\lambda)k'_{\mu}=0.$
The terms proportional to  $k'_{\mu}$
modify the
Green function $\tau^{\mu}$ in (2.1), but for the on shell 
amplitude 
they can be ignored. In addition, due to $k'_{\mu}k'^{\mu}\equiv {k'}^2=0$
 the terms proportional to  $k'_{\mu}$  in (2.12b) can also be omitted
in the Ward-Takahashi identity (2.14b).

Current conservation (2.7)  and (2.14b)  are 
written for the  longitudinal part of 
the total $\gamma'\pi'N'-\pi N$ amplitude. In particular,
the set of the diagrams which form   
the first term of the right side in  (2.2b) with 
${{\partial}/{\partial z^{\mu}} }\ {\cal J}^{\mu}(z)=0$
are not included in (2.7).
Therefore, the modified Ward-Takahashi identity  (2.7) presents a necessary 
condition of current 
conservation which  contains only a longitudinal part of the 
external particle radiation amplitudes.

One can use the transverse part of the total $\pi N$ bremsstrahlung amplitude
in order to complete ${\cal E}_{\gamma'\pi' N'-\pi N}^{\mu}(N)$ (2.12b)
up to the external particle radiation amplitude with the anomalous magnetic 
moment of the external nucleons. For this aim one can pick out the related  
transverse terms with $\sigma^{\mu\nu}k'_{\nu}$ which generate the following 
redefinitions of $\mu_{N'}=1$ and $\mu_N=1$
with the corresponding anomalous magnetic moments of the 
nucleon{\footnotemark}.
\footnotetext{
Keeping  the identity (2.14b) one can also reproduce   
the full electromagnetic form factors of the nucleon ${\em F}_1(t)$ and
${\em F}_2(t)$ in ${\cal E}_{\gamma'\pi' N'-\pi N}^{\mu}(N)$ (2.12b),
where $t$ denotes the corresponding four momentum transfer.
Thus, if one picks out the terms   
$ {\overline u}({\bf p'}_N)\Bigl[(2p'_N+k')^{\mu}
-i{\widetilde \mu}_{N'}\sigma^{\mu\nu}k'_{\nu}\Bigr]{\em F}_2(t')
{\em F}_1^{-1}(t')u({\bf p_N'+k'}){\cal T}_{N'} $
and
${\cal T}_{N} {\overline u}({\bf p_N-k'})\Bigl[
(2p_N-k')^{\mu}-i{\widetilde \mu_N}\sigma^{\mu\nu}k'_{\nu}{\em F}_2(t)
{\em F}_1^{-1}(t)\Bigr]u({\bf p}_N) $ from the transverse part of the
full $\pi N$ radiation amplitude, one obtains the full 
$\gamma NN$ vertices in ${\cal E}_{\gamma'\pi' N'-\pi N}^{\mu}(N)$ (2.12b)
with the redefined $\pi N$ amplitudes (2.9a,c) 
${\cal T}_{N'}\Longrightarrow {\em F}_1^{-1}(t'){\cal T}_{N'}$ and 
${\cal T}_{N}\Longrightarrow {\em F}_1^{-1}(t){\cal T}_{N}$. 

Another way to take into account the full electromagnetic 
form factors of the external nucleons is to use 
the modified complete set of the intermediate Dirac spinors 
$u({\bf p_N'+k'}){\overline u}({\bf p_N'+k'})
\Longrightarrow$
\newline $u_{m_N\to\sqrt{s_N'} }({\bf p_N'+k'})
{\overline u}_{m_N\to\sqrt{s_N'} }({\bf p_N'+k'});\ 
u({\bf p_N-k'}){\overline u}({\bf p_N-k'}) u_{m_N\to\sqrt{s_N} }({\bf p_N-k'})
{\overline u}_{m_N\to\sqrt{s_N} }({\bf p_N-k'}),$
where $m_N$ is replaced by
$s_N'=\Bigl(\sqrt{m_N^2+({\bf p_N'})^2}+k'\Bigr)^2-({\bf p'_N+k'})^2$
or $s_N=\Bigl(\sqrt{m_N^2+({\bf p_N})^2}-k'\Bigr)^2-({\bf p_N- k'})^2$.
Then we obtain the $\gamma' NN$ vertex between the one nucleon states
with the four moments $p_{N'}+k'\to p_{N'}$ or $p_{N}-k'\to p_{N}$.
In this formulation $t=t'={k'}^2=0$ and 
only the threshold values of the
electromagnetic form factors for the external nucleons appear.
               
The large four momentum transfer of the external 
nucleons is not important for the determination of the 
electromagnetic moments of the $\Delta$'s. 
Therefore, we do not include them in the following text.}
This procedure  implies taking into account  the loop
corrections  of the $\gamma NN$ vertex. These corrections
 reproduces the anomalous magnetic moments within 
 the minimal electromagnetic coupling scheme
(see \cite{BD1} consideration of  (10.81)).
Then we obtain{\footnotemark} 
\footnotetext{
The anomalous magnetic moment of the $\Delta$ appears in the 
$\Delta-\gamma\Delta$ vertex at 
$\sigma_{\mu\nu}{k'}^{\nu}_{\Delta}$ with ${k'}^{\nu}_{\Delta}\ne 
{k'}^{\nu}$, i.e., the diagram in Fig. 2B with the anomalous magnetic moment 
of $\Delta$ is not included in the transverse part of the 
$\pi N$ bremsstrahlung amplitude.}

$${\cal E}_{\gamma'\pi' N'-\pi N}^{\mu}(N)=
-\Biggl[(2{p'}_{\pi}+k')^{\mu}{ {e_{\pi'}}
\over{ {2p'_{\pi}k' }} }
<out;{\bf p'}_{N}|j_{\pi'}(0)|{\bf p}_{\pi}{\bf p}_N;in>
$$
$$+{{
{\overline u}({\bf p'}_N)\Bigl[(2p'_N+k')^{\mu}
-i\mu_{N'}\sigma^{\mu\nu}k'_{\nu}\Bigr] }\over{ 2p'_Nk'} }
u({\bf p_N'+k'}){\overline u}({\bf p_N'+k'})
 e_{N'}<out;{\bf p'}_{\pi}|J(0)|{\bf p}_{\pi}{\bf p}_N;in>
$$
$$
-e_N<out;{\bf p'}_{\pi}{\bf p'}_{N}|{\overline J}(0)|{\bf p}_{\pi};in>
u({\bf p_N-k'}){\overline u}({\bf p_N-k'})
{ {\Bigl[(2p_N-k')^{\mu}-i\mu_{N}\sigma^{\mu\nu}k'_{\nu}
\Bigr]u({\bf p}_N) }\over{2p_Nk'} }$$
$$-<out;{\bf p'}_{\pi}{\bf p'}_{N}|j_{\pi}(0)|{\bf p}_{N};in>
{{e_{\pi}}\over{2p_{\pi}k'} }
(2p_{\pi}-k')^{\mu} 
\Biggr],\eqno(2.15)$$

The external particle $\pi N$ radiation amplitude (2.15) have the fixed
transverse terms $i\mu_{N}\sigma^{\mu\nu}k'_{\nu}$ in the 
vertex functions of the external nucleons.
The  full $\gamma NN$ vertices are necessary
for  the realistic calculations of the $\pi N$ radiation reactions.
In particular, expression (2.15) automatically satisfies  
the low energy photon theorem \cite{Low}-\cite{Lin}.

Relation (2.14b) represents the modified Ward-Takahashi identity
in the three-dimensional time-ordered form. This identity establish 
a relationship between the external particle 
$\pi N$ bremsstrahlung amplitude (2.15) and the
off mass shell elastic $\pi N$ scattering amplitudes (2.9a,b,c,d).
In order to satisfy current conservation (2.14b)
it is necessary to find the internal particle radiation diagrams
whose four divergence reproduces ${\cal B}_{\pi' N'-\pi  N}(N)$.
The special case of this problem for the (3/2,3/2) partial 
$\pi N$ amplitudes is considered in the next Section.

\vspace{0.25cm}

\begin{center}
   {\bf 3. Internal and external  particle radiation 
parts of the $\pi N$ bremsstrahlung amplitude.}
\end{center}

\vspace{0.15cm}

In this Section we show, that
the Ward-Takahashi identity (2.14b) after  decompositions 
of $({\cal E})^{\mu}_{\gamma'\pi'N'-\pi N}(N)$ (2.15) 
and ${\cal B}_{\pi'N'-\pi N}(N)$ (2.12a) reduces to a special identity for  
the double $\Delta$ exchange amplitude which has the same structure
as on mass shell $\Delta$ radiation diagram in Fig. 2B.  
For this aim we separate the longitudinal part 
of the external particle radiation amplitude and 
isolate
the $\Delta$ resonance parts  of the
off shell $\pi N$ amplitudes (2.9a,b,c,d).
 The symbolic representation of this procedure is given by
the chain of transformations
$$({\cal E})^{\mu}_{\gamma'\pi'N'-\pi N}(N)\Longrightarrow
({\cal E_L})^{\mu}_{\gamma'\pi'N'-\pi N}\Longrightarrow
({\cal E_L}^{3/2})^{\mu}_{\gamma'\pi'N'-\pi N}(\Delta-pole)\eqno(3.1a)$$
and 
$${\cal B}_{\pi'N'-\pi N}(N)\Longrightarrow
({\cal B}^{3/2})_{\pi'N'-\pi N}(\Delta-pole),\eqno(3.1b)$$ 
where the lower index ${\cal L}$
denotes the longitudinal part of  the amplitude
$({\cal E})^{\mu}_{\gamma'\pi'N'-\pi N}$, the upper index 
$3/2$ corresponds to the resonance spin-isospin $3/2$ state of the 
 $\pi N$ amplitudes. The argument $(\Delta-pole)$ indicates the
$\Delta$-pole part of the $\pi N$ amplitudes (2.9a,b,c,d).

Afterwards, using the sum of the $\Delta$-pole terms in the different 
off shell $\pi N$ amplitudes in 
$({\cal E_L}^{3/2})^{\mu}_{\gamma'\pi'N'-\pi N}(\Delta-pole)$ and 
$({\cal B}^{3/2})_{\pi'N'-\pi N}(\Delta-pole)$,
one can separate the  double $\Delta$ exchange 
Ward-Takahashi identities for the double
$\Delta$ exchange amplitudes
$({\cal E_L}^{3/2})^{\mu}_{\gamma'\pi'N'-\pi N}(\Delta\Delta)$ and 
$({\cal B}^{3/2})_{\pi'N'-\pi N}(\Delta\Delta)$. Moreover, after an
algebraic transformation of the Ward-Takahashi identity for the
double $\Delta$  exchange amplitudes
one obtains an independent
identity for the  amplitude which has the same structure  
as the internal $\Delta$ radiation amplitude in Fig. 2B.
These decompositions form the following chain of transformations

$$({\cal E_L}^{3/2})^{\mu}_{\gamma'\pi'N'-\pi N}(\Delta-pole)
\Longrightarrow
({\cal E_L}^{3/2})^{\mu}_{\gamma'\pi'N'-\pi N}(\Delta\Delta)
\Longrightarrow
({\cal E_L}^{3/2})^{\mu}_{\gamma'\pi'N'-\pi N}(\Delta-\gamma\Delta)
\eqno(3.1c)$$

$$({\cal B}^{3/2})_{\pi'N'-\pi N}(\Delta-pole)\Longrightarrow
({\cal B}^{3/2})_{\pi'N'-\pi N}(\Delta\Delta)\Longrightarrow
{\cal B}^{3/2}_{\pi'N'-\pi N}(\Delta-\gamma\Delta)\eqno(3.1d)$$

The algebraic decompositions (3.1a,b,c,d) 
of the Ward-Takahashi identity (2.14b)
are detailed  in Appendix A. The resulting identity is

$${k'}_{\mu} 
\Biggl[<out;{\bf p'}_{N}{\bf p'}_{\pi}|
{\cal J}^{\mu}(0)|{\bf p}_{\pi}{\bf p}_N;in>
\Biggr]^{Projection\ on\ spin\ 3/2\ particle\  states}
_{2\Delta\ exchange\ with\ a\ \Delta-\gamma'\Delta'\ vertex}=$$
$$k'_{\mu}
({\cal E_L}^{3/2})_{\gamma'\pi' N'-\pi N}^{\mu}({\Delta}-\gamma\Delta)
+{\cal B}^{3/2}_{\pi' N'-\pi N}(\Delta-\gamma\Delta)=0,\eqno(3.2)$$

where

$$({\cal E_L}^{3/2})_{\gamma'\pi' N'-\pi N}^{\mu}({\Delta}-\gamma\Delta)=
{{<{\bf p'}_N,{\bf p'}_{\pi}|{\sl g}_{\pi' N'-\Delta'}|{\bf P'}_{\Delta}>}
\over{{p'}_{\pi}^{o}+{p'}_{N}^{o}-{P'}^o_{\Delta}(s')}}$$
$$ \Biggl\{
{\overline u}^b({\bf P'}_{\Delta})g_{bc}\Bigl[
(P_{\Delta}+P'_{\Delta})^{\mu}{\sc V}_E-
i\sigma^{\mu\nu}{k'_{\Delta}}_{\nu}{\sc V}_H
\Bigr]u^c({\bf P}_{\Delta})\Biggr\}
{{<{\bf P}_{\Delta}|{\sl g}_{\Delta-\pi N}|{\bf p}_N,{\bf p}_{\pi}>}
\over{{p}_{\pi}^{o}+{p}_{N}^{o}-{P}^o_{\Delta}(s) }}
,\eqno(3.3a)$$

$${\cal B}^{3/2}_{\pi' N'-\pi N}(\Delta-\gamma\Delta)=
{{<{\bf p'}_N,{\bf p'}_{\pi}|{\sl g}_{\pi' N'-\Delta'}|{\bf P'}_{\Delta}>}
\over{{p'}_{\pi}^{o}+{p'}_{N}^{o}-{P'}^o_{\Delta}(s')}}$$
$$ \Biggl\{
{\overline u}^b({\bf P'}_{\Delta})g_{bc}\Bigl[
k'_o(P_{\Delta}+P'_{\Delta})^{o}{\sc V}_E-
ik'_{\mu}\sigma^{\mu o}{k'_{\Delta}}_{o}{\sc V}_H
\Bigr]u^c({\bf P}_{\Delta})\Biggr\}
{{<{\bf P}_{\Delta}|{\sl g}_{\Delta-\pi N}|{\bf p}_N,{\bf p}_{\pi}>}
\over{{p}_{\pi}^{o}+{p}_{N}^{o}-{P}^o_{\Delta}(s) }}
,\eqno(3.3b)$$
where $P=p_{\pi}+p_N$ and $P'=p'_{\pi}+p'_N$  are the four moments of
the $\pi N$ system in the initial and final states, 
$P_{\Delta}$ and $P'_{\Delta}$ denote the four moments of the 
$\Delta$

$$P_{\Delta}\equiv\Bigl(P^o_{\Delta}(s),{\bf P}_{\Delta}\Bigr)=\Bigl(\sqrt{
(M_{\Delta}(s)-{{i\Gamma_{\Delta}(s)}\over 2})^2
+{\bf P}_{\Delta}^2},{\bf P}_{\Delta}\Bigr)
;\ \ \ \ \ \ \ \ {\bf P}_{\Delta}={\bf P}={\bf p}_N+{\bf p}_{\pi}\eqno(3.4a)$$
$$P'_{\Delta}\equiv\Bigl({P'}^o_{\Delta}(s'),{\bf P'}_{\Delta}\Bigr)=\Bigl(
\sqrt{(M_{\Delta}(s')-{{i\Gamma_{\Delta}(s')}\over 2})^2
+{\bf P'}_{\Delta}^2 },{\bf P'}_{\Delta}\Bigr);\ \ \ \ \ \ 
 {\bf P'}_{\Delta}={\bf P'}={\bf p'}_N+{\bf p'}_{\pi}.\eqno(3.4b)$$

  We shall use  two  models of the   
$\Delta$ mass \cite{Ann,MF,NP}
${\sf m}_{\Delta}(s)=M_{\Delta}(s)-i/2\Gamma_{\Delta}(s)$:

\begin{itemize}

\item[$1.$]
A model with the fixed mass of the intermediate  $\Delta$ resonance

$${\sf m}_{\Delta}=M_{\Delta}-{i\over 2}\Gamma_{\Delta}
=1232MeV - {i\over  2}120MeV,\eqno(3.4c)$$

\item[$2.$]
and a more general model with an $s$-dependent 
mass ${\sf m}_{\Delta}(s)$

$${\sf m}_{\Delta}(s)=M_{\Delta}(s)-{i\over 2}\Gamma_{\Delta}(s),\eqno(3.4d)$$ 
where ${\sf m}_{\Delta}(s=M_{\Delta}^2)={\sf m}_{\Delta}$.

\end{itemize}

$P_{\Delta}$ and $P'_{\Delta}$ are  on mass shell 
four moments
because $P_{\Delta}^2={\sf m}_{\Delta}^2(s)$  and 
${P'_{\Delta} }^2={\sf m}_{\Delta}^2(s')$.    
$u^b({\bf P_{\Delta} })$ denotes the Rarita-Schwinger
spinor of the free spin  $3/2$
particle with the complex mass ${\sf m}_{\Delta}(s)$.  

${\sc V}_E$ and ${\sc V}_H$ in (3.3a,b) are defined through the 
$\Delta$-pole residues ${\cal R}_{N'}$,
${\cal R}_{\pi'}$, ${\cal R}_{N}$ and ${\cal R}_{\pi}$
of the off shell $\pi N$ amplitudes  in 
equations (A.9a,b,c,d), (A.15a,b) and (A.17a,b) of Appendix A. 
The  $\Delta-\pi N$ and $\pi N-\Delta$ vertices ${\sl g}_{\pi'
  N'-\Delta}$
and ${\sl g}_{\Delta-\pi N}$   
are defined as

$$<{\bf p'}_N,{\bf p'}_{\pi}|{\sl g}_{\pi' N'-\Delta}|{\bf P'}_{\Delta}>=
{\rm g}_{\pi' N'-\Delta'}(s',k')
{\overline u}({\bf p'}_N)i\gamma_5
{ {(p'_N)_a}\over{ |{\bf p'}_N| }}
u^a({\bf P'}_{\Delta}),
\eqno(3.5a)$$

$$<{\bf P}_{\Delta}|{\sl g}_{\Delta-\pi N}|{\bf p}_N,{\bf p}_{\pi}>=
{\rm g}_{\Delta-\pi N}(s)
{\overline u}^d({\bf P}_{\Delta})
{ {(p_N)_d}\over{ |{\bf p}_N| }}
i\gamma_5 u({\bf p}_N).
\eqno(3.5b)$$

The longitudinal part of the external particle radiation amplitude
with the $\Delta$ intermediate states 
$({\cal E_L}^{3/2})_{\gamma'\pi' N'-\pi N}^{\mu}({\Delta}-\gamma\Delta)$
 (3.3a) has the same form as the internal $\Delta$ radiation
amplitude ${\cal I}_{\gamma'\pi' N'-\pi N}^{\mu}(\Delta-\gamma\Delta)$
in Fig. 2B

$${\cal I}_{\gamma'\pi' N'-\pi N}^{\mu}(\Delta-\gamma\Delta)=-
{{<{\bf p'}_N,{\bf p'}_{\pi}|{\sl g}_{\pi' N'-\Delta'}|{\bf P'}_{\Delta}>}
\over{{p'}_{\pi}^{o}+{p'}_{N}^{o}-{P'}^o_{\Delta}(s')}}$$
$$<{\bf P'}_{\Delta},{\sf m}_{\Delta}(s')|{\cal J}^{\mu}(0)
|{\bf P}_{\Delta},{\sf m}_{\Delta}(s)>
{{<{\bf P}_{\Delta}|{\sl g}_{\Delta-\pi N}|{\bf p}_N,{\bf p}_{\pi}>}
\over{{p}_{\pi}^{o}+{p}_{N}^{o}-{P}^o_{\Delta}(s) }},
\eqno(3.6)$$

where  the details of the $\Delta-\gamma\Delta$ vertex   
$<{\bf P'}_{\Delta},{\sf m}_{\Delta}(s')|{\cal J}^{\mu}(0)
|{\bf P}_{\Delta},{\sf m}_{\Delta}(s)>$
with on mass shell $\Delta$'s are given in Appendix B.
In particular, for the low energy photons we have

$$
<{\bf P'}_{\Delta},{\sf m}_{\Delta}(s')|J_{\mu}(0)|{\bf P}_{\Delta},
{\sf m}_{\Delta}(s)>=$$
$${\overline u}^{\sigma}({\bf P'}_{\Delta})g_{\rho\sigma}\biggl[ 
{{(P_{\Delta}+P'_{\Delta})_{\mu}}\over{2M_{\Delta}} }
G_{C0}({k'}_{\Delta}^2,s,s')- 
{{i\sigma_{\mu\nu}{k'}_{\Delta}^{\nu}}\over{2M_{\Delta} }}
G_{M1}({k'}_{\Delta}^2,s,s')
\biggr]u^{\rho}({\bf P}_{\Delta})\eqno(3.7)$$
where 
$G_C$ and $G_{M1}$ denote the electric and 
magnetic dipole form factors of the $\Delta$'s.

The unambiguous construction of 
the $\Delta$ radiation amplitude 
${\cal I}_{\gamma'\pi' N'-\pi N}^{\mu}({\Delta}-\gamma{\Delta})$
(3.6) is given in Appendix C following our previous papers 
\cite{Ann,MF}. Thus    
${\cal I}_{\gamma'\pi' N'-\pi N}^{\mu}({\Delta}-\gamma{\Delta})$
in Fig. 2B can be determined as 
a projection of the complete internal particle
radiation amplitude

$$
{\cal I}_{\gamma'\pi' N'-\pi N}^{\mu}({\Delta}-\gamma{\Delta})=
\Biggl[ {\cal I}_{\gamma'\pi' N'-\pi N}
^{\mu}\Biggr]^{Projection\ on\ spin\ 3/2\ particle\  states}
_{2\Delta\ exchange\ with\ a\ \Delta-\gamma'\Delta'\ vertex},$$

Consequently, expressions (3.3a) and (3.6) determine the complete
projection of the $\pi N$ radiation amplitude, and we have

$${k'}_{\mu} 
\Biggl[<out;{\bf p'}_{N}{\bf p'}_{\pi}|
{\cal J}^{\mu}(0)|{\bf p}_{\pi}{\bf p}_N;in>
\Biggr]^{Projection\ on\ spin\ 3/2\ particle\  states}
_{2\Delta\ exchange\ with\ a\ \Delta-\gamma'\Delta'\ vertex}=$$
$$k'_{\mu}\biggl[ 
({\cal E_L}^{3/2})_{\gamma'\pi' N'-\pi N}^{\mu}({\Delta}-\gamma\Delta)
+{\cal I}_{\gamma'\pi' N'-\pi N}^{\mu}(\Delta-\gamma\Delta)\biggr]=0.
\eqno(3.8)$$

Combining  (3.2) and (3.8) we obtain

$$k'_{\mu}
({\cal E_L}^{3/2})_{\gamma'\pi' N'-\pi N}^{\mu}({\Delta}-\gamma\Delta)
=-k'_{\mu}{\cal I}_{\gamma'\pi' N'-\pi N}^{\mu}(\Delta-\gamma\Delta)
=-{\cal B}^{3/2}_{\pi' N'-\pi N}(\Delta-\gamma\Delta)\eqno(3.9)$$

Conditions  (3.2) and (3.9)
present the four-divergence of the amplitudes  \newline
$({\cal E_L}^{3/2})_{\gamma'\pi' N'-\pi N}^{\mu}({\Delta}-\gamma\Delta)$ 
(3.3a) and 
$ {\cal I}_{\gamma'\pi'N'-\pi N}^{\mu}(\Delta-\gamma\Delta)$ (3.6) 
with the same double $\Delta$ poles and  
the  corresponding $\Delta-\gamma\Delta$ vertex. There are no
other amplitudes with the same analytical structure.
Moreover, all gauge terms $A_{\mu}$
with  the separate current conservation condition 
$k'_{\mu}A^{\mu}=0$ are included in the transverse part of the external 
particle radiation amplitude
$({\cal E_{TR}})_{\gamma'\pi'N'-\pi N}^{\mu}$ (A.4b)
and other transverse parts of the total amplitude
which corresponds to the 
first term in (2.2b) with  $\partial/\partial z^{\mu}\ {\cal J}_{\mu}(z)$. 
Therefore, current conservation (3.2) and (3.9) are fulfilled if

$$({\cal E_L}^{3/2})_{\gamma'\pi' N'-\pi N}^{\mu}({\Delta}-\gamma\Delta)=
-{\cal I}_{\gamma'\pi' N'-\pi N}^{\mu}(\Delta-\gamma\Delta)
\eqno(3.10)$$

This equation coincides with  the final equation (1.5) and 
allows  to determinate the connection between
the form factors  of the $\Delta-\gamma\Delta$ vertices in 
 ${\cal I}_{\gamma'\pi' N'-\pi N}^{\mu}(\Delta-\gamma\Delta)$ 
(3.6) and  the analogical  form factors in
$({\cal E_L}^{3/2})_{\gamma'\pi' N'-\pi N}^{\mu}({\Delta}-\gamma\Delta)$ 
(3.3a).

 Using the condition  (3.10) one easily  gets

$${k'_{\Delta}}_{\mu} 
\biggl[ ({\cal E_L}^{3/2})_{\gamma'\pi' N'-\pi N}^{\mu}({\Delta}-\gamma\Delta) 
+{\cal I}^{\mu}_{\gamma'\pi' N'-\pi N}({\Delta}-\gamma\Delta)
\biggr]=0\eqno(3.11)$$

which immediately gives

$$G_{C0}({k'}_{\Delta}^2,s,s')=-
2M_{\Delta}{\sc V}_E(s'{\bf P'}_{\Delta};s{\bf P}_{\Delta})
\eqno(3.12a)$$

Combining this equation with  (3.24) we obtain 

$$G_{M1}({k'}_{\Delta}^2,s,s')=
-2M_{\Delta}{\sc V}_H(s'{\bf P'}_{\Delta};s{\bf P}_{\Delta})
.\eqno(3.12b)$$

Equations (3.12a,b) determine  $G_{C0}({k'}_{\Delta}^2,s,s')$ and
$G_{M1}({k'}_{\Delta}^2,s,s')$ via the residues of the $\pi N$
amplitudes ${\cal  R}$ (A.9a,b,c,d)
which yield $V_E$ and $V_H$ in (A.17a,b) and 
${\cal V}^{(+)}_E$ and ${\cal V}^{(+)}_H$ in (A.15a,b). 

The threshold values of  (3.12a,b) give a relations between
$e_{\Delta}$, $\mu_{\Delta}$ and  
${\sc V}_E^{k'=0}$, ${\sc V}_H^{k'=0}$ respectively

$$e_{\Delta}=
G_{C0}({k'}_{\Delta}^2=0,s'=M_{\Delta}^2,s=M_{\Delta}^2)=
-\biggl[2M_{\Delta}{\sc V}_E\biggr]^{k'=0}
_{\sqrt{s'}=\sqrt{s}=M_{\Delta}}\eqno(3.13a)$$

and

$$\mu_{\Delta}=G_{M1}({k'_{\Delta}}^2=0,s'=M_{\Delta}^2,s=M_{\Delta}^2)
=-\biggl[{2M_{\Delta}\sc V}_H\biggr]^{k'=0}
_{\sqrt{s'}=\sqrt{s}=M_{\Delta}}\eqno(3.13b)$$

These conditions  determine the relations between $e_{\Delta}$,
$\mu_{\Delta}$  and residues  of the $\pi N$ amplitudes  ${\cal R}$
 (A.9a,b,c,d).

$$e_{\Delta}=-\Biggl[{\cal N}(s)
\Bigl[{\rm g}_{\pi' N'-\Delta'}(s',k')\Bigr]^{-1}
\biggl(e_N{{{\cal R}_{N'}+{\cal R}_N}\over 2}
+e_{\pi}{{{\cal R}_{\pi'}+{\cal R}_{\pi}}\over 2}\biggr)
\Bigl[{\rm g}_{\Delta-\pi N}(s)\Bigr]^{-1}
\Biggr]^{k'=0}
_{\sqrt{s'}=\sqrt{s}=M_{\Delta}},
\eqno(3.14a)$$
where 
${\cal N}(s)=1/(d{\sqrt{s}}/dk')-d{P}^o_{\Delta}(s)/d{\sqrt{s}} $ and

$$\mu_{\Delta}=-\Biggl[
{\cal N}(s)
\Bigl[{\rm g}_{\pi' N'-\Delta'}(s',k')\Bigr]^{-1}
\biggl(\mu_N {{{\cal R}_{N'}+{\cal R}_N}\over 2} \biggr)
\Bigl[{\rm g}_{\Delta-\pi N}(s)\Bigr]^{-1}
\Biggr]^{k'=0}
_{\sqrt{s'}=\sqrt{s}=M_{\Delta}}.
\eqno(3.14b)$$

The similarity of the conditions (3.14a) and (3.14b) allows  
to determine $\mu_{\Delta}$ using (3.14a) as 
a normalization condition.
For this aim we consider (3.14b) separately for the 
$\pi^+n\to \gamma'{\pi'}^+n'$ and $\pi^op\to \gamma'{\pi'}^op'$
reactions. 
For the $\pi^+n\to \gamma'{\pi'}^+n'$ reaction  
(3.14a) generates the independent normalization condition

$$
1=
-\Biggl[{\cal N}(s)
\Bigl[{\rm g}_{\pi' N'-\Delta'}(s',k')\Bigr]^{-1}
{{{\cal R}_{\pi'}+{\cal R}_{\pi}}\over 2}
\Bigl[{\rm g}_{\Delta-\pi N}(s)\Bigr]^{-1}
\Biggr]^{k'=0}
_{\sqrt{s'}=\sqrt{s}=M_{\Delta}}
\eqno(3.15)$$

and for the $\pi^op\to \gamma'{\pi'}^op'$ reaction we get 

$$1=
-\Biggl[{\cal N}(s)
\Bigl[{\rm g}_{\pi' N'-\Delta'}(s',k')\Bigr]^{-1}
{{{\cal R}_{p'}+{\cal R}_p}\over 2}
\Bigl[{\rm g}_{\Delta-\pi N}(s)\Bigr]^{-1}.
\Biggr]^{k'=0}
_{\sqrt{s'}=\sqrt{s}=M_{\Delta}}
\eqno(3.16)$$

Expressions (3.15) and (3.16) 
are  the normalization conditions
for the  residues 
${\cal R}_{\pi}$ and ${\cal R}_p$ 
(A.9a,b,c,d) of  the $\pi N$ matrices 
at the $\Delta$ resonance pole position.  
They show the  dependence of 
${\cal R}_{\pi}$ and ${\cal R}_p$ (A.9a,b,c,d) on the 
 $\Delta$ mass    
${\sf m}_{\Delta}(s)$ (3.4c,d).
Therefore, ${\rm g}_{\Delta-\pi N}(s)$
and ${\rm g}_{\pi' N'-\Delta'}(s',k')$ form factors
must also include a dependence on a ${\sf m}_{\Delta}(s)$.

The right side of (3.16)
differs from the right side of (3.14b) only by the factor $\mu_N$.
Therefore, substituting (3.16) into (3.14b) we obtain

$$\mu_{\Delta^+}=
\mu_p{ {M_{\Delta} }\over {m_p} },\eqno(3.17)$$
where 
the factor ${ {M_{\Delta} }/ {m_p} }$ arises because of the different
units for $\mu_{\Delta^+}$ and $\mu_p$.

The magnetic dipole moment of $\Delta^{++}$ can be determined from 
the relationship between 
$\mu_{\Delta}$ and $G_{M1}({k'_{\Delta}}^2=0,s'=M_{\Delta}^2,s=M_{\Delta}^2)$
(3.13b) and (3.14b). The difference between 
$({\cal R}_{N'}+{\cal R}_N)/2$ in (3.16) for the $\pi^o p\to\gamma'\pi'^o p'$
and $\pi^+ p\to\gamma'\pi'^+ p'$
reactions is in the isospin factors of the corresponding 
$\pi N$ amplitudes.
Using the isotospin 
symmetry between the $\pi^o p\to \pi^o p$ and $\pi^+ p\to \pi^+ p$
 amplitudes  we get  
$$\mu_{\Delta^{++}}={3\over 2}\mu_{\Delta^+}={3\over 2}
\mu_p{ {M_{\Delta} }\over {m_p} }
\eqno(3.18)$$

One cannot use directly
the $\pi^o n\to\gamma' {\pi^o}' n'$
and  $\pi^- n\to\gamma' {\pi^-}' n'$
reactions for determination of 
the magnetic moments of $\Delta^{o}$ and $\Delta^{-}$, because
the equal-time commutator (2.3a) is zero in this case. 
This problem is considered in the next part of the present paper.

 The modified Ward-Takahashi identity  (3.9)
requires  equality and cancellation of 
the internal $\Delta$ radiation amplitude in Fig. 2B
and the corresponding part 
of the external particle radiation amplitude
according to relation (3.10). 
This cancellation is the result of current conservation  
for the $\pi N$ radiation amplitude and the
special sum of the off shell $\pi N$ amplitudes 
in $({\cal E_L}^{3/2})_{\gamma'\pi' N'-\pi N}(\Delta-\gamma\Delta)$
(3.3a), which have the same analytical structure  
as  the internal $\Delta$ radiation amplitude (3.6) in Fig. 2B. 
Therefore, the amplitude 
(3.6) in Fig. 2B  cancels exactly  with    
$({\cal E_L}^{3/2})_{\gamma'\pi' N'-\pi N}^{\mu}({\Delta}-\gamma\Delta)$ 
(3.3a).
Consequently,  the internal $\Delta$ radiation
diagram in Fig. 2B is screened by
the appropriate part of the external particle radiation diagrams.
Generally, screening  is built into  the initial 
Ward-Takahashi identity (2.7),  where
${\cal B}_{\pi' N'-\pi N}$ must be compensated by the 
internal particle radiation diagrams.
In other words, the screening corresponds to equality and
cancellation of special parts of the internal  
and external particle  radiation terms in the total $\pi N$
bremsstrahlung amplitude.


\vspace{0.25cm}

\begin{center}
                  {\bf 4. Conclusion}
\end{center}

\vspace{0.25cm}

In the present paper  $\Delta$'s are 
considered as resonances of the $\pi N$
system which  generate  appropriate $\Delta$-poles  
in the off mass shell $\pi N$ amplitudes 
(A.9a,b,c,d). The sums of the  corresponding  residues 
determine the $\Delta$ form factors $G_C$ and $G_{M1}$ in (3.14a,b).
Thus current conservation  (3.2) makes it possible to 
determine  $G_C$ and $G_{M1}$ only
using the dynamical 
information about the residues of the off shell $\pi N$ amplitudes
(2.9a,b,c,d). The threshold values  of $G_C$ and $G_{M1}$ define the 
magnetic dipole moments of $\Delta^+$ and $\Delta^{++}$ via the anomalous
 magnetic moment of the proton. The difference between $\mu_{\Delta}$ and 
$\mu_p$ is formed by different units in the  
$\Delta-\gamma\Delta$ and $p-\gamma p$ electromagnetic vertex functions.
Another dynamical input for reproduction of the magnetic dipole moment of
 $\Delta^+$ and $\Delta^{++}$ is the anomalous magnetic moment of
the proton, which requires  loop corrections for the 
$\gamma NN$ vertices and the corresponding redefinition
of the external particle radiation amplitude (2.12b) by  expression 
(2.15). 
This redefinition does not change the initial 
current conservation (2.7) and 
 is necessary for reproduction of the realistic results 
for the $\pi N$ bremsstrahlung reactions.

The present investigation of the $\pi N$ radiation is  
based on current conservation for the on  shell amplitudes 
(2.7). 
From the general point of view only the sum of the external and 
internal particle radiation parts of the full bremsstrahlung amplitude 
satisfies current conservation. The modified Ward-Takahashi identity (2.7)
specifies this statement for the special form of the external particle 
radiation 
amplitude ${\cal E}_{\gamma'\pi' N'-\pi N}^{\mu}$  
and the appropriate  sum of the off shell  $\pi N$ scattering 
amplitudes ${\cal B}_{\pi' N'-\pi N}$.
In particular,
${\cal E}_{\gamma'\pi' N'-\pi N}^{\mu}$ contains only the diagrams which are 
responsible for the infrared behavior of the $\pi N$ radiation amplitude. 
The sums of the $\Delta$-pole $\pi N$ amplitudes in 
the longitudinal part of  ${\cal E}_{\gamma'\pi' N'-\pi N}^{\mu}$
reproduce the double $\Delta$ exchange poles.

The model-independent properties of current conservation  (2.7) can 
be generalized for any amplitude of an arbitrary reaction
$a+b\longrightarrow \gamma'+f_1+...+f_n$ $(n=1,2,...)$.
Current conservation  requires the  existence of the internal particle 
radiation amplitude ${\cal I}_{\gamma'f_1..f_n-a b}^{\mu}$ which 
satisfies the relation
 ${k'}_{\mu}{\cal I}_{\gamma'f_1...f_n-a b}^{\mu}
=-{k'}_{\mu}{\cal E}_{\gamma'f_1...f_n-a b}^{\mu}
={\cal B}_{f_1...f_n-a b}$,
where ${\cal E}_{\gamma'f_1...f_n-a b}^{\mu}$ is the external particle
radiation amplitude.
Therefore, the appropriate parts of  
${\cal E}_{\gamma'f_1..f_n-a b}^{\mu}$ and 
${\cal I}_{\gamma'f_1...f_n-a b}^{\mu}$ have a different sign and they must be
subtracted from each other. Consequently, we have a screening  of the 
internal particle 
radiation amplitudes by the external one-particle radiation terms.
In the limit $k'\to 0$ our approach 
 exactly reproduces
the low energy photon theorems for the bremsstrahlung reactions.

As an example of the screening 
the identity and cancellation of the double $\Delta$ 
exchange amplitude in  Fig. 2B 
and the longitudinal part of the external particle
radiation amplitude is demonstrated in (3.10).
This cancellation allows  to determine 
the  magnetic dipole moments
$\mu_{\Delta^+}=G_{M1}(0)={ {M_{\Delta} }\over {m_N} }\mu_p$ and 
$\mu_{\Delta^{++}}={3\over 2}\mu_{\Delta^+}=5.46\mu_B$ or 
$\mu_{\Delta^{++}}/\mu_p\sim 1.95$
 of the $\Delta^+$ and $\Delta^{++}$ resonances. Our result
for $\mu_{\Delta^{++}}$   
roughly agrees with the prediction of the naive $SU(6)$ quark 
model for $\mu_{\Delta^{++}}=2\mu_p=5.58\mu_B$ \cite{Beg,Georgi},
with the nonrelativistic potential model \cite{Boss}
$\mu_{\Delta^{++}}=4.52\pm0.95\mu_B$
and with  extraction of $\mu_{\Delta^{++}}$ from the 
 experimental $\pi ^{+}p\to\gamma\pi^{+}p$ cross section
within the low energy photon approach
 $\mu_{\Delta^{++}}=3.6\pm2.0\mu_B$ \cite{Musa}, 
$\mu_{\Delta^{++}}=5.6\pm2.1\mu_B$ \cite{Pascual} and 
$\mu_{\Delta^{++}}=4.7-6.9\mu_B$\cite{Nefkens}. 
Our result is larger than  the predictions of the modified 
$SU(6)$ models \cite{Brown,Pais} and 
the low energy photon approximation 
$\mu_{\Delta^{++}}=3.7\sim 4.9 \mu_B$ \cite{Lin}.
On the other hand,
our result is smaller than the values obtained within the 
effective meson-nucleon Lagrangian  
$\mu_{\Delta^{++}}=6.1\pm 0.5 \mu_B$ \cite{Castro},
 in the effective quark model $\mu_{\Delta^{++}}=6.17\mu_B$ \cite{Franklin},
 in the modified bag model $\mu_{\Delta^{++}}=6.54\mu_B$ \cite{Kriv},
and in the constituent quark model \cite{Buch97}.

The resulting magnetic dipole moments of $\Delta$'s 
 obtained in various theoretical models differ.
Moreover, the results obtained using the same Low theorem approach
for soft  photons also differ.
This difference can be explained with the various recipes 
for the construction of the bremsstrahlung amplitude in the 
low energy photon limit $k'\to 0$.
These  ambiguities are 
listed in \cite{Ding}. Our formulation is free  
off these ambiguities.

\vspace{0.1cm}

\begin{center}
{\bf Table  1.\ \ \ }{\em Magnetic moments of $\Delta^+$ and $\Delta^{++}$
 in units of the nuclear magneton $\mu_B={e/{2m_N}}$. 
 The upper index ${\bf ^*}$ at the reference indicates  the theoretical model  
  which is used to fit of the experimental data and to extract the 
magnetic moment   $\mu_{\Delta}$. }   
\end{center}

\begin{center}

\hfill{\small

\begin{tabular}{|l|l|l|l|l|l|l|l|}\hline\hline
{\small Model}   &  {\small This}            &  {\small $SU(6)$}  
                 &{\small Potential and}     &  {\small Skyrme }
                 &{\small Low ener. phot.}   &  {\small Eff. $\pi N$ }
            &                    
\\
            &      {\small  work }             & {\small and Bag }
            &      {\small K-matr. app.}       & 
            &      {\small theorem  }          & {\small Lagran. }      
            &      {\small quark }         

\\    \hline

                       &                             & 2.79 \cite{Beg,Georgi}
                       &                             & 
                       &                             &
                       &3.49\cite{Buch97}         
                                 
\\

${ \mu_{\Delta^{+}}}$ & 3.66                   &  2.13\cite{Pais} 
                &                               &  2.0-3.0\cite{Acu98}
                &                               &
                       & 2.85\cite{Kim98}

\\
                       &                             & 2.20-2.45\cite{Brown}
                       &                             &
                       &                             &
                       &  2.3-2.7\cite{Lin98}         

\\                     &                             &3.27\cite{Kriv} 
                       &                             &
                       &                             &
                       &  2.79\cite{Franklin}

\\    \hline

                       &                             & 5.58 \cite{Beg,Georgi}
                       &6.9-9.7\cite{Heller}$^*$     & 
                       &3.6$\pm$2.0\cite{Musa}$^*$   &
                       &6.98\cite{Buch97}         
                                 
\\

${ \mu_{\Delta^{++}}}$ & 5.49                    &  4.25\cite{Pais} 
                & 4.52$\pm$0.95\cite{Boss}$^*$     &  4.2-7.4\cite{Acu98}
                & 5.6$\pm$2.1\cite{Pascual}$^*$  &6.1$\pm$0.5\cite{Castro}$^*$
                       & 5.33\cite{Kim98}

\\
                       &                           & 4.41-4.89\cite{Brown}
                       &  5.6-7.5\cite{Wittman}$^*$  &
                       &  4.7-6.9\cite{Nefkens}$^*$  &
                       &  5.1-5.4\cite{Lin98}         

\\                     &                             &6.54\cite{Kriv} 
                       &                             &
                       &  3.7-4.9\cite{Lin}$^*$      &
                       &  6.17\cite{Franklin}

\\ \hline \hline

\end{tabular}

}

\end{center}

\vspace{0.1cm}

The numerical values of the magnetic moments of
the $\Delta^+$ and $\Delta^{++}$ resonances are given in Table 1. 
In a number of approaches the magnetic moment of  $\Delta$
is treated as an adjustable parameter. The corresponding results obtained
 from the experimental cross sections of 
the $\pi^+p\to\gamma\pi^+p$ reaction are indicated 
in Table 1 with the  upper index $^*$.
It must be emphasized that only our approach and the naive $SU(6)$ quark model
give an analytical form for $\mu_{\Delta^{+}}$ and $\mu_{\Delta^{++}}$.
But our result for $\mu_{\Delta^{+}}$ is $M_{\Delta}/m_p\sim 1.31$ times
larger as  
$\mu_{\Delta^{+}}=\mu_{p}=2.79\mu_B$ in refs. \cite{Beg,Franklin}.

The $SU(6)$ models \cite{Beg,Georgi,Pais} and their 
bag model modifications require proportionality between  the charge and 
 the magnetic dipole moment $\mu_{\Delta}=e_{\Delta} \mu_p$
of the $\Delta$ resonance. Therefore, 
$\mu_{\Delta^+}=1/2\ \mu_{\Delta^{++}}$ in  \cite{Beg,Georgi,Pais,Brown,Kriv}.
This property is preserved  in the constituent quark model \cite{Buch97}.
But it is broken in the Skyrme model \cite{Acu98},
 chiral quark model \cite{Lin98}, chiral quark-soliton 
model \cite{Kim98},
and effective quark model \cite{Franklin}. 
Our result for the ratio $\mu_{\Delta^{++}}/ \mu_{\Delta^{+}}$
is determined by the isospin factors of the $\pi^+ p$ and $\pi^o p$
elastic scattering amplitudes. In addition, we  take into account 
the  difference between units of $\mu_{\Delta}$ and $\mu_p$ in
the $\Delta-\gamma\Delta$ and $\gamma NN$ 
vertices.  This difference generates the factor 
$M_{\Delta}/m_N$. Therefore 
the present value $\mu_{\Delta^{+}}=3.64\mu_B$ is larger than
other predictions.

Our  approach is based on usual local quantum field theory\cite{BD2,IZ}. 
This approach is not dependent on the form of the Lagrangian.
Moreover,  we have not used a special representation of the $\pi N$
amplitude and the $\Delta$ propagator.
Therefore, the suggested relations between $\mu_{\Delta}$
and the anomalous magnetic moment of the proton are model independent.
 But the present 
field-theoretical formulation does not include
the quark degrees of freedom. 
The generalization of the present formulation 
based on the field-theoretical approach with the
quark-gluon degrees of freedom will be
given in the following paper.

\vspace{0.5cm}

We thank P. Minkowski for his interest in this paper and 
M. I. Potapov  for his help in preparation of  this manuscript.


\begin{center}
                  {{\bf Appendix A:} Projections on the intermediate  
$\Delta$ states}
\end{center}
\medskip

\vspace{0.25cm}

In this section  
a set of transformations (3.1a,c) of the 
external particle radiation amplitude 
$({\cal E})^{\mu}_{\gamma'\pi'N'-\pi N}(N)$(2.15) and the
corresponding Ward-Takahashi identity (2.14b) is performed.
The resulting condition (3.2) as well as other intermediate
current conservation conditions are obtained on the basis  of the algebraic 
identity  

$$k'_{\mu}\Bigl[(P+P')^{\mu}-i\sigma^{\mu\nu}k'_{\nu}\Bigr]=s-s'.
\eqno(A.1)$$

The decompositions (3.1a,b,c,d) of current conservation  
are detailed  in the following subsections.

\begin{center}

 {\em A. Decomposition over the transverse and longitudinal parts of the 
external particle radiation amplitude (2.15), i.e.,
$({\cal E})^{\mu}_{\gamma'\pi'N'-\pi N}(N)\Longrightarrow$
$({\cal E_L})^{\mu}_{\gamma'\pi'N'-\pi N}(N)$.}

\end{center}
\vspace{0.15cm}

In order to separate the transverse  part of
 ${\cal E}_{\gamma'\pi'  N'-\pi N}^{\mu}(N)$ (2.15) (Fig. 1) 
it is convenient to introduce the total and relative moments
$$P=p_{\pi}+p_N;\ \ \ 
p={{\alpha_{\pi}p_N-\alpha_{N}p_{\pi}}\over{\alpha_{\pi}+\alpha_N}};\ \ \  
p_N={{\alpha_{N}P}\over{\alpha_{\pi}+\alpha_N}}+p,\ \ \
p_{\pi}={{\alpha_{\pi}P}\over{\alpha_{\pi}+\alpha_N}}-p,\eqno(A.2a)$$

$$P'=p'_{\pi}+p'_N;\ \ \ 
p'={{\alpha'_{\pi}p'_N-\alpha'_{N}p'_{\pi}}
\over{\alpha'_{\pi}+\alpha'_N}};\ \ \  
p'_N={{\alpha'_{N}P'}\over{\alpha'_{\pi}+\alpha'_N}}+p',\ \ \
p'_{\pi}={{\alpha'_{\pi}P'}\over{\alpha'_{\pi}+\alpha'_N}}-p',\eqno(A.2b)$$

where

$$\alpha_{N}=k'_{\nu}p_{N}^{\nu},\ \ \
\alpha_{\pi}=k'_{\nu}p_{\pi}^{\nu};\ \ \
\alpha'_{N}=k'_{\nu}{p'_{N}}^{\nu},\ \ \
\alpha'_{\pi}=k'_{\nu}{p'_{\pi}}^{\nu}.\eqno(A.2c)$$

The relative moments $p$ and $p'$ are transverse to $k'_{\mu}$

$$k'_{\nu}{p}^{\nu}=0;\ \ \ k'_{\nu}{p'}^{\nu}=0.\eqno(A.2d)$$

Now one can  separate 
the transverse part $({\cal E_{TR}})_{\gamma'\pi' N'-\pi N}^{\mu}$
from  ${\cal E}_{\gamma'\pi' N'-\pi N}^{\mu}(N)$  (2.15) as

$${\cal E}_{\gamma'\pi' N'-\pi N}^{\mu}(N)=
({\sf E_L})_{\gamma'\pi' N'-\pi N}^{\mu}
+({\cal E_{TR}})_{\gamma'\pi' N'-\pi N}^{\mu},
\eqno(A.3)$$
where ${k'}_{\mu}({\sf E_L})_{\gamma'\pi' N'-\pi N}^{\mu}
=-{\cal B}_{\pi' N'-\pi N}(N)$,       
${k'}_{\mu}({\cal E_{TR}})_{\gamma'\pi' N'-\pi N}^{\mu}=0$
 and

$$({\sf E_L})_{\gamma'\pi' N'-\pi N}^{\mu}=
-{1\over{s-s'}}\Biggl[ 
{\overline u}({\bf p'_N})\Bigl[2e_{N'}{P'}^{\mu}
-i\mu_{N'}\sigma^{\mu\nu}k'_{\nu}\Bigr]u({\bf p'_N+k'})
{\overline u}({\bf p'_N+k'})<out;{\bf p'}_{\pi}|J(0)|{\bf p}_{\pi}{\bf p}_N;in>
$$
$$-<out;{\bf p'}_{\pi}{\bf p'}_{N}|{\overline J}(0)|{\bf p}_{\pi};in>
u({\bf p_N-k'}){\overline u}({\bf p_N-k'})\Bigl[2e_NP^{\mu}
-i\mu_N\sigma^{\mu\nu}k'_{\nu}
\Bigr]u({\bf p}_N)\Biggr]$$
$$-{1\over{s-s'}}\Biggl[2e_{\pi'}{P'}^{\mu}
<out;{\bf p'}_{N}|j_{\pi'}(0)|{\bf p}_{\pi}{\bf p}_N;in>
-2e_{\pi}P^{\mu}
<out;{\bf p'}_{\pi}{\bf p'}_{N}|j_{\pi}(0)|{\bf p}_{N};in>
\Biggr]\eqno(A.4a)$$

$$({\cal E_{TR}})_{\gamma'\pi' N'-\pi N}^{\mu}=
-\Biggl[ e_{N'}{{{p'}^{\mu}}\over{\alpha'_N}}
{\overline u}({\bf p'}_N)u({\bf p'_N+k'})
{\overline u}({\bf p'_N+k'})<out;{\bf p'}_{\pi}|J(0)|{\bf p}_{\pi}{\bf p}_N;in>
$$
$$-e_{\pi'}{{{p'}^{\mu}}\over{\alpha'_{\pi} }}
<out;{\bf p'}_{N}|j_{\pi'}(0)|{\bf p}_{\pi}{\bf p}_N;in>
+e_{\pi}{{{p}^{\mu}}\over{\alpha_{\pi} } }
<out;{\bf p'}_{\pi}{\bf p'}_{N}|j_{\pi}(0)|{\bf p}_{N};in>$$
$$-e_{N}{{{p}^{\mu}}\over{\alpha_N}}
<out;{\bf p'}_{\pi}{\bf p'}_{N}|{\overline J}(0)|{\bf p}_{\pi};in>
u({\bf p_N-k'}){\overline u}({\bf p_N-k'})u({\bf p}_N)
\Biggr]\eqno(A.4b)$$

Other non-longitudinal terms can be obtained using
 new total four moments $P_{\pm}$ 

$$P_{\pm}={1\over2}(P\pm P'),\ \ where \ P=P_+ +P_-;\ \ \ P'=P_+-P_-
\ \ \ and\ \ \ {P_-}^{\mu}={1\over 2}{k'}^{\mu}.\eqno(A.5)$$

This allows to separate of  the term 
${\cal K}_{\gamma'\pi' N'-\pi N}^{\mu}$  proportional to $k'_{\mu}$

$$({\sf E_L})_{\gamma'\pi' N'-\pi N}^{\mu}=
  ({\cal E_L})_{\gamma'\pi' N'-\pi N}^{\mu}+
  {\cal K}_{\gamma'\pi' N'-\pi N}^{\mu},\eqno(A.6)$$

where

$$({\cal E_L})_{\gamma'\pi' N'-\pi N}^{\mu}=
-{1\over{s-s'}}
\Biggl[ {\overline u}({\bf p'_N})\Bigl[e_{N'}(P+P')^{\mu}
-i\mu_{N'}\sigma^{\mu\nu}k'_{\nu}\Bigr]u({\bf p'_N+k'}) 
{\overline u}({\bf p'_N+k'})<out;{\bf p'}_{\pi}|J(0)|{\bf p}_{\pi}{\bf p}_N;in>
$$
$$-<out;{\bf p'}_{\pi}{\bf p'}_{N}|{\overline J}(0)|{\bf p}_{\pi};in>
u({\bf p_N-k'}){\overline u}({\bf p_N-k'})\Bigl[e_N(P'+P)^{\mu}
-i\mu_N\sigma^{\mu\nu}k'_{\nu}
\Bigr]u({\bf p}_N)\Biggr]$$
$$-{1\over{s-s'}}\Biggl[e_{\pi'}{(P+P')}^{\mu}
<out;{\bf p'}_{N}|j_{\pi'}(0)|{\bf p}_{\pi}{\bf p}_N;in>
-e_{\pi}{(P+P')}^{\mu}
<out;{\bf p'}_{\pi}{\bf p'}_{N}|j_{\pi}(0)|{\bf p}_{N};in>
\Biggr],\eqno(A.7a)$$

$${\cal K}_{\gamma'\pi' N'-\pi N}^{\mu}={{{ k'}^{\mu}}\over{s-s'}}\Biggl[ 
 {\overline u}({\bf p'}_N)
u({\bf p_N'+k'}){\overline u}({\bf p_N'+k'})
e_{N'}<out;{\bf p'}_{\pi}|J(0)|{\bf p}_{\pi}{\bf p}_N;in>$$
$$+e_{\pi'}<out;{\bf p'}_{N}|j_{\pi'}(0)|{\bf p}_{\pi}{\bf p}_N;in>$$
$$+e_N<out;{\bf p'}_{\pi}{\bf p'}_{N}|{\overline J}(0)|{\bf p}_{\pi};in>
u({\bf p_N-k'}){\overline u}({\bf p_N-k'})
u({\bf p}_N)+e_{\pi}
<out;{\bf p'}_{\pi}{\bf p'}_{N}|j_{\pi}(0)|{\bf p}_{N};in>
\Biggr],\eqno(A.7b)$$

It is easy to see that 
${k'}_{\mu}{\cal E}_{\gamma'\pi' N'-\pi N}^{\mu}(N)=
{k'}_{\mu}({\cal E_L})_{\gamma'\pi' N'-\pi N}^{\mu}
=-{\cal B}_{\pi' N'-\pi N}(N)$.
The resulting expression
$({\cal E_L})_{\gamma'\pi' N'-\pi N}^{\mu}$ (3.7a) differs
from  ${\cal E}_{\gamma'\pi' N'-\pi N}^{\mu}(N)$ (2.15)  
by the $\gamma N N$ and $\gamma \pi\pi$ vertices 
which have unified factors  
$(P+P')^{\mu}-i\mu_{N'(N)}\sigma^{\mu\nu}k'_{\nu}$
and   $(P+P')^{\mu}$.


\begin{center}

{\em B. 
Projection on the  $\Delta$-pole terms 
$({\cal E_L})^{\mu}_{\gamma'\pi'N'-\pi N}(N)\Longrightarrow$
$({\cal E_L}^{3/2})^{\mu}_{\gamma'\pi'N'-\pi N}(\Delta-pole)$
 and 
${\cal B}_{\pi'N'-\pi N}(N)\Longrightarrow$
${\cal B}^{3/2}_{\pi'N'-\pi N}(\Delta-pole)$}.

\end{center}


In order  to separate  the  $\Delta$-pole parts 
in $({\cal E_L})_{\gamma'\pi' N'-\pi N}^{\mu}$ (A.7a) we shall use 
a projection of  
the $\gamma N-N$ and $\gamma\pi-\pi$ vertex 
on the spin $3/2$ intermediate states

$$(p'_N{\bf .}p_N)\biggl[{\overline u}({\bf p'_N})\gamma_5
\Bigl[e_{N'}{(P+P')}^{\mu}-i\mu_{N'}\sigma^{\mu\nu}k'_{\nu}\Bigr]\gamma_5
u({\bf p'_N+k'})\biggr]^{Projection\ on\ spin\ 3/2\ states}=$$
$$\biggl( {\overline u}({\bf p'_N})\gamma_5{p'_N}_au^a({\bf P'})\Biggl\{
{\overline u}^b({\bf P'})g_{bc}\biggl[e_{N'}{(P+P')}^{\mu}
-i\mu_{N'}\sigma^{\mu\nu}k'_{\nu}\biggr]u^c({\bf P})\Biggr\}
{\overline u}^d({\bf P})(p_N)_d\gamma_5u({\bf p_N})\biggl)$$
$${\overline  u}({\bf p_N})
u({\bf p'_N+k'}), \eqno(A.8a)$$

$$(p'_N{\bf .}p_N)\biggl[
e_{\pi'}{(P+P')}^{\mu}
\biggr]^{Projection\ on\ spin\ 3/2\ states}=
\biggl( {\overline u}({\bf p'_N})\gamma_5{p'_N}_au^a({\bf P'})$$
$$\Bigl[{\overline u}^b({\bf P'})g_{bc}e_{\pi'}{(P+P')}^{\mu}
u^c({\bf P})\Bigr]
{\overline u}^d({\bf P})(p_N)_d\gamma_5u({\bf p_N})\biggl)
{\overline  u}({\bf p_N})
u({\bf p'_N}). \eqno(A.8b)$$

where $(p'_N{\bf .}p_N)=(p'_{N})_{\sigma} (p_N)^{\sigma}$
and we  omit the spin index $S=\pm 1/2,\pm 3/2$ of
the Rarita-Schwinger spinor  $u^a({\bf P})$ with the  
mass $m_{D_{3/2}}^2=P_o^2-{\bf P}^2$.  
The matrix element (A.8a) corresponds   
to the transitions $\pi N\to D_{3/2}\to
\gamma'D_{3/2}'\to\pi'N'$ with the intermediate spin $3/2$ particles $D_{3/2}$
and $D_{3/2}'$.{\footnotemark}

The common factor  
$\biggl( {\overline u}({\bf p'_N})\gamma_5{p'_N}_au^a({\bf P'})$
 $\Bigl[{\overline u}^b({\bf P'})g_{bc}(...)^{\mu}
u^c({\bf P})\Bigr]
{\overline u}^d({\bf P})(p_N)_d\gamma_5u({\bf p_N})\biggl)$
{\footnotemark}
in  (A.8a,b) generates the following redefinition of (A.7a)

$$\biggl[({\cal E_L})_{\gamma'\pi' N'-\pi N}^{\mu}\biggr]^{Projection\ on
\ spin\ 3/2\ states}
\equiv({\cal E_L}^{3/2})_{\gamma'\pi' N'-\pi N}^{\mu}=$$
$${{ (p'_N)_{a} (p_N)_d {(P+P')}^{\mu}}
\over{|{\bf p'}_N||{\bf p}_N|(s-s')}}
{\overline u}({\bf p'_N})i\gamma_5u^a({\bf P'})
\Bigl\{
{\overline u}^b({\bf P'})g_{bc}u^c({\bf P})\Bigr\}
{\overline u}^d({\bf P})i\gamma_5u({\bf p_N})$$
$$\Biggl[{{|{\bf p'}_N||{\bf p}_N|}\over{  (p'_N{\bf .}p_N)}}
\biggl({\overline u}({\bf p_N})
u({\bf p_N'+k'}){\overline u}({\bf p_N'+k'})
e_{N'}<out;{\bf p'}_{\pi}|J(0)|{\bf p}_{\pi}{\bf p}_N;in>$$
$$-e_{N}<out;{\bf p'}_{\pi}{\bf p'}_{N}|{\overline J}(0)|{\bf p}_{\pi};in>
u({\bf p_N-k'}){\overline u}({\bf p_N-k'})u({\bf p'_N})$$
$$+{\overline u}({\bf p_N})u({\bf p'_N})
e_{\pi'}<out;{\bf p'}_{N}|j_{\pi'}(0)|{\bf p}_{\pi}{\bf p}_N;in>$$
$$-e_{\pi}<out;{\bf p'}_{\pi}{\bf p'}_{N}|j_{\pi}(0)|{\bf p}_{N};in>
{\overline u}({\bf p_N})u({\bf p'_N})\biggr)
\Biggr]^{Projection\ on\ spin\ 3/2\ states}$$
$$+{{ (p'_N)_{a} (p_N)_d }\over{ 
 |{\bf p'}_N||{\bf p}_N|(s-s')}}
{\overline u}({\bf p'_N})i\gamma_5u^a({\bf P'})\Bigl\{
{\overline u}^b({\bf P'})g_{bc}
(-i\sigma^{\mu\nu}k'_{\nu})u^c({\bf P})\Bigr\}
{\overline u}^d({\bf P})i\gamma_5u({\bf p_N})$$
$$\Biggl[ 
{{|{\bf p'}_N||{\bf p}_N|}\over{  (p'_N{\bf .}p_N)}}
\biggl( {\overline u}({\bf p_N})
u({\bf p_N'+k'}){\overline u}({\bf p_N'+k'})
\mu_{N'}<out;{\bf p'}_{\pi}|J(0)|{\bf p}_{\pi}{\bf p}_N;in>$$
$$-\mu_{N}<out;{\bf p'}_{\pi}{\bf p'}_{N}|{\overline J}(0)|{\bf p}_{\pi};in>
u({\bf p_N-k'}){\overline u}({\bf p_N-k'})u({\bf p'_N})
\biggr)\Biggr]^{Projection\ on\ spin\ 3/2\ states}
.\eqno(A.8c)$$

\footnotetext{
Using the completeness conditions of the spin $3/2$ 
 functions $u^{a}({\bf P},S)$ 
\cite{Nie,W,Bammer} and
$v^{a}({\bf P},S)$ of the free particle and antiparticle states

$$\sum_{S=-3/2}^{3/2}\Biggl(
 u^{a}({\bf P},S){\overline u}^{b}({\bf P},S)+
{ {\gamma_{\sigma}P^{\sigma}+s^{1/2} }\over {2s^{1/2}} }
\biggl\{\biggl[projections\ on\ spin\ 1/2\ states\biggr]^{ab}\biggr\}\Biggr)+$$
$$\sum_{S=-3/2}^{3/2}\Biggl(
 v^{a}({\bf P},S){\overline v}^{b}({\bf P},S)+
{ -{\gamma_{\sigma}P^{\sigma}+s^{1/2} }\over {2s^{1/2}} }
\biggl\{\biggl[projections\ on\ spin\ 1/2\ states
\biggr]^{ab}\biggr\}^{\ast}\Biggr)
=g^{ab}$$

one can rewrite the $\gamma N-N$ vertex as

$$
g_{bc}{\overline u}({\bf p'_N})\gamma_5g^{ab}
\Bigl[e_{N'}{(P+P')}^{\mu}-i\mu_{N'}\sigma^{\mu\nu}k'_{\nu}\Bigr]
g^{cd}\gamma_5 u({\bf p'_N+k'})=
{\overline u}({\bf p'_N})\gamma_5
\sum_{S=-3/2}^{3/2}\Biggl(
 u^{a}({\bf P},S){\overline u}^{b}({\bf P},S)+
v^{a}({\bf P},S){\overline v}^{b}({\bf P},S)+...\Biggr)$$
$$g_{bc}\Bigl[e_{N'}{(P+P')}^{\mu}-i\mu_{N'}\sigma^{\mu\nu}k'_{\nu}\Bigr]
\sum_{S'=-3/2}^{3/2}\Biggl(
 u^{c}({\bf P},S){\overline u}^{d}({\bf P},S)+
v^{c}({\bf P},S){\overline v}^{d}({\bf P},S)+...\Biggr)
\gamma_5u({\bf p'_N+k'}).$$

In the $\Delta$ resonance region one can take into account only
the spin $3/2$ intermediate states. 
In addition,
other degrees of freedom with antiparticle and  spin $1/2$ intermediate states
form the independent Ward-Takahashi identities. 
Then for the final nucleon radiation term  we obtain

$$\biggl[(p'_{N})_{a}(p_N)^{a}{\overline u}({\bf p'_N})\gamma_5
\Bigl[e_{N'}{(P+P')}^{\mu}-i\mu_{N'}\sigma^{\mu\nu}k'_{\nu}\Bigr]\gamma_5
u({\bf p'_N+k'})\biggr]^{Projection\ on\ spin\ 3/2\ states}=$$
$$\sum_{S,S'=-3/2}^{3/2} 
{\overline u}({\bf p'_N})\gamma_5{p'_N}_au^a({\bf P'},S')\Biggl\{
{\overline u}^b({\bf P'},S')g_{bc}\biggl[e_{N'}{(P+P')}^{\mu}
-i\mu_{N'}\sigma^{\mu\nu}k'_{\nu}\biggr]u^c({\bf P},S)\Biggr\}
{\overline u}^d({\bf P},S)(p_N)_d\gamma_5
u({\bf p'_N+k'}).$$

In the same way for the final pion radiation term  we get 
 
$$<out;{\bf p'}_{N}|j_{\pi'}(0)|{\bf p}_{\pi}{\bf p}_N;in>\Longrightarrow
{\overline u}({\bf p'}_N)\gamma_5^2 u({\bf p}_N)
{\overline u}({\bf p}_N)u({\bf p'}_N)
<out;{\bf p'}_{N}|j_{\pi'}(0)|{\bf p}_{\pi}{\bf p}_N;in>\Longrightarrow
{1\over{(p'_N{\bf .}p_N) }}$$
$$\Biggl[
{\overline u}({\bf p'}_N)\gamma_5(p'_N)_au^a({\bf P'})\biggl\{
{\overline u}^b({\bf P'})g_{bc}u^c({\bf P})\biggr\}
{\overline u}^d({\bf P})(p_N)_d\gamma_5u({\bf p}_N)\Biggr]
{\overline u}({\bf p}_N)u({\bf p'}_N)
<out;{\bf p'}_{N}|j_{\pi'}(0)|{\bf p}_{\pi}{\bf p}_N;in>.$$

Hereafter  the 
spin indexes $S$ and $S'$ are omitted for the sake of simplicity.
}

The $\pi N$ amplitudes (2.9a,b,c,d) consist of the  resonant
and non-resonant parts {\footnotemark}

$$\biggl[
{{|{\bf p'}_N||{\bf p}_N|}\over{  (p'_N{\bf .}p_N)}}
{\overline u}({\bf p_N}){{ (p'_N+k')_{\sigma}\gamma^{\sigma}+m_N}\over {2m_N}}
{\overline u}({\bf p'}_N)
<out;{\bf p'}_{\pi}|J(0)|{\bf p}_{\pi}{\bf p}_N;in>)
\biggr]^{Projection\ on\ spin\ 3/2\ states}=$$
$${{ {\cal R}_{N'}(s'{\bf P'}_{\Delta};s{\bf P}_{\Delta})
}\over{{p}_{\pi}^{o}+{p}_{N}^{o}-P^o_{\Delta}(s)}}+
{r}_{N'}(s'{\bf P'}_{\Delta};s{\bf P}_{\Delta})
;\eqno(A.9a)$$

$$\biggl[
{{|{\bf p'}_N||{\bf p}_N|}\over{  (p'_N{\bf .}p_N)}}
{\overline u}({\bf p_N})u({\bf p'_N})
<out;{\bf p'}_{N}|j_{\pi'}(0)|{\bf p}_{\pi}{\bf p}_N;in>
\biggr]^{Projection\ on\ spin\ 3/2\ states}=$$
$${{ {\cal R}_{\pi'}(s'{\bf P'}_{\Delta};s{\bf P}_{\Delta})}\over{
{p}_{\pi}^{o}+{p}_{N}^{o}-
{P}^o_{\Delta}(s) }}+{r}_{\pi'}(s'{\bf P'}_{\Delta};s{\bf  P}_{\Delta})
;\eqno(A.9b)$$

$$\bigg[
{{|{\bf p'}_N||{\bf p}_N|}\over{  (p'_N{\bf .}p_N)}}
<out;{\bf p'}_{\pi}{\bf p'}_{N}|{\overline J}(0)|{\bf p}_{\pi};in>
u({\bf p}_N){{ (p_N-k')_{\sigma}\gamma^{\sigma}+m_N}\over {2m_N}}
u({\bf p'_N})
\biggr]^{Projection\ on\ spin\ 3/2\ states}=$$
$${{ {\cal R}_{N}(s'{\bf P'}_{\Delta};s{\bf P}_{\Delta}) }\over{
{p'}_{\pi}^{o}+{p'_{N}}^{o}-
{P'}^o_{\Delta}(s')} }+{r}_{N}(s'{\bf P'}_{\Delta};s{\bf P}_{\Delta});
\eqno(A.9c)$$
$$\biggl[
{{|{\bf p'}_N||{\bf p}_N|}\over{  (p'_N{\bf .}p_N)}}
<out;{\bf p'}_{\pi}{\bf p'}_{N}|j_{\pi}(0)|{\bf p}_{N};in>
{\overline u}({\bf p_N})u({\bf p'_N})
\biggr]^{Projection\ on\ spin\ 3/2\ states}=$$
$${{ {\cal R}_{\pi}(s'{\bf P'}_{\Delta};s{\bf P}_{\Delta}) }\over{
{p'}_{\pi}^{o}+{p'}_{N}^{o}-{P'}^o_{\Delta}(s') }}
+{r}_{\pi}(s'{\bf P'}_{\Delta};s{\bf P}_{\Delta}).\eqno(A.9d)$$

The linear relativistic  $\Delta$ propagator 
in  (A.9a,b,c,d)  
is most  similar to non-relativistic quantum mechanical 
 $\Delta$ propagator.

\footnotetext{ 
For the on energy shell $\pi N$ and $\pi'N'$ states with
$s=s'$, i.e., $k'=0$, the operator

$$ {\cal Q}({\bf p'}_N,{\bf p}_N,{\bf P'}_{\Delta},{\bf P}_{\Delta})=
{1\over{(p'_N{\bf .}p_N) }}
{\overline u}({\bf p'}_N)i\gamma_5({{p'}_N})_a u^a({\bf P'}_{\Delta})
{\overline u}^d({\bf P}_{\Delta})({{p}_N})_d i\gamma_5 u({\bf p}_N)$$
is transformed into the projection operator 
on the $\pi N$ state with 
the orbital momentum $L=1$ and the total momentum $J=3/2$ 
${\cal P}_1^{3/2}({\bf p'}_N,{\bf p}_N)$  \cite{MF}

$${\cal P}_1^{3/2}({\bf p'}_N,{\bf p}_N)=
{{6m_N}\over{4\pi{\bf p}{\bf p'}(m_N+\sqrt{m_N^2+{\bf p}^2})}}
{\overline u}({\bf p'}_N)i\gamma_5{{p'}_N}_a u^a({\bf P'}_{\Delta})
{\overline u}^d({\bf P}_{\Delta}){{p}_N}_d i\gamma_5 u({\bf p}_N).$$

Therefore, we have

$$\Biggl[{\cal Q}({\bf p'}_N,{\bf p}_N,{\bf P'}_{\Delta},{\bf P}_{\Delta})
\Biggr]^{|{\bf k'}|=0}
= {{ 4\pi{\bf p}{\bf p'}(m_N+\sqrt{m_N^2+{\bf p}^2} ) }
\over{6m_N (p'_N{\bf .}p_N) }}
{\cal P}_1^{3/2}({\bf p'}_N,{\bf p}_N)$$
}

\footnotetext{
Equations (A.9a,b,c,d) are also valid 
in the models, where the $\Delta$'s are considered as the 
intermediate one-particle states  which decays into the  
asymptotic $\pi N$ states
(see for example  ref. \cite{Thomas}).
The ${\cal R}$ and $r$ functions and the $\Delta-\pi N$
vertices in this case are defined in the one-particle 
approach. Nevertheless, intermediate one-particle $\Delta$'s are 
not equivalent to real particle degrees of freedom which have  
appropriate one-particle asymptotic states.
Therefore, the introduction of the effective $\pi N\Delta$ 
Lagrangian with the 
one-particle Heisenberg operator of  $\Delta$ 
requires an additional assumption.}

Projection on the spin $3/2$ states and separation of the 
$\Delta$ pole terms modifies Ward-Takahashi identity (A.7c)
as 

$$
\Biggl[{k'}_{\mu}
<out;{\bf p'}_{N}{\bf p'}_{\pi}|{\cal J}^{\mu}(0)|{\bf p}_{\pi}{\bf p}_N;in>
\Biggr]
^{Projection\ on\ spin\ 3/2\ particle\  states}
={k'}_{\mu}({\cal E_L}^{3/2})_{\gamma'\pi' N'-\pi N}^{\mu}
+{\cal B}^{3/2}_{\pi' N'-\pi N}$$
$$={k'}_{\mu}
\Bigl(({\cal E_L}^{3/2})_{\gamma'\pi' N'-\pi N}^{\mu}({\Delta}-pole)
+({\cal E_L}^{3/2})_{\gamma'\pi' N'-\pi N}^{\mu}(non-pole)\Bigr)
+{\cal B}^{3/2}_{\pi' N'-\pi N}({\Delta}-pole)
+{\cal B}^{3/2}_{\pi' N'-\pi N}(non-pole)=0.\eqno(A.10a)$$

With identity (A.1), it is easy to see, that the $\Delta$-pole and 
$non-pole$ parts of $({\cal E_L}^{3/2})_{\gamma'\pi' N'-\pi N}^{\mu}$
separately satisfy the independent Ward-Takahashi identities. 
In particular,

$$\Biggl[{k'}_{\mu}
<out;{\bf p'}_{N}{\bf p'}_{\pi}|{\cal J}^{\mu}(0)|{\bf p}_{\pi}{\bf p}_N;in>
\Biggr]
^{Projection\ on\ spin\ 3/2\ particle\  states}
_{\Delta-pole}$$
$$
{k'}_{\mu}({\cal E_L}^{3/2})_{\gamma'\pi' N'-\pi N}^{\mu}({\Delta}-pole)
+{\cal B}^{3/2}_{\pi' N'-\pi N}({\Delta}-pole)=0,\eqno(A.10b)$$
where 
$$({\cal E_L}^{3/2})_{\gamma'\pi' N'-\pi N}^{\mu}({\Delta}-pole)=
 {1\over{ |{\bf p'}_N||{\bf p}_N| }}
{\overline u}({\bf p'_N})(p'_N)_{a}i\gamma_5u^a({\bf P'})$$
$$\Biggl\{
{\overline u}^b({\bf P'})g_{bc}\Bigl[
(P+P')^{\mu}{\cal V}_E-i\sigma^{\mu\nu}k'_{\nu}{\cal V}_H
\Bigr]u^c({\bf P})\Biggr\}
{\overline u}^d({\bf P})(p_N)_d i\gamma_5u({\bf p_N}),\eqno(A.10c)$$

$${\cal B}^{3/2}_{\pi' N'-\pi N}({\Delta}-pole)=
{{ (p'_N)_{a} (p_N)_d }\over{ 
|{\bf p'}_N||{\bf p}_N|}}
{\overline u}({\bf p'_N})i\gamma_5u^a({\bf P'})\Bigl\{
{\overline u}^b({\bf P'})g_{bc}u^c({\bf P})\Bigr\}
{\overline u}^d({\bf P})i\gamma_5u({\bf p_N})$$
$$\Biggl[
{{ {\cal R}_{N'}(s'{\bf P'}_{\Delta};s{\bf P}_{\Delta})}\over{
{p}_{\pi}^{o}+{p}_{N}^{o}-{P}^o_{\Delta}(s) }}
+{{ {\cal R}_{\pi'}(s'{\bf P'}_{\Delta};s{\bf P}_{\Delta})}\over{
{p}_{\pi}^{o}+{p}_{N}^{o}-
{P}^o_{\Delta}(s) }}-
{{ {\cal R}_{N}(s'{\bf P'}_{\Delta};s{\bf P}_{\Delta}) }\over{
{p'}_{\pi}^{o}+{p'}_{N}^{o}-{P'}^o_{\Delta}(s') }}
-{{ {\cal R}_{\pi}(s'{\bf P'}_{\Delta};s{\bf P}_{\Delta}) }\over{
{p'}_{\pi}^{o}+{p'}_{N}^{o}-{P'}^o_{\Delta}(s') }}
\Biggr],\eqno(A.10d)$$

where 

$${\cal V}_E=
{{ e_{N'}{\cal R}_{N'}}\over{(s-s')
\Bigl(
{p}_{\pi}^{o}+{p}_{N}^{o}-{P}^o_{\Delta}(s)\Bigr)}}
+{{e_{\pi'} {\cal R}_{\pi'}}
\over{(s-s')
\Bigl(
{p}_{\pi}^{o}+{p}_{N}^{o}-{P}^o_{\Delta}(s)\Bigr) }}$$
$$-{{e_{N}{\cal R}_{N}}
\over{(s-s')\Bigl(
{p'}_{\pi}^{o}+{p'}_{N}^{o}-{P'}^o_{\Delta}(s')\Bigr) }}
-{{e_{\pi} {\cal R}_{\pi} }\over{
(s-s')
\Bigl({p'}_{\pi}^{o}+{p'}_{N}^{o}-{P'}^o_{\Delta}(s')
\Bigr) }},
\eqno(A.11a)$$

$${\cal V}_H=
{{ \mu_{N'}{\cal R}_{N'}}\over{(s-s')
\Bigl( {p}_{\pi}^{o}+{p}_{N}^{o}-{P}^o_{\Delta}(s)\Bigr) }}
-{{\mu_{N}{\cal R}_{N}}
\over{(s-s')
\Bigl({p'}_{\pi}^{o}+{p'}_{N}^{o}-{P'}^o_{\Delta}(s')
\Bigr) }},\eqno(A.11b)$$

Hereafter we omit the variables of ${\cal R}$ 
functions  for the sake of simplicity .

The resulting expressions (A.10c,d) have the 
common factors 
${\overline u}({\bf p'_N})(p'_N)_{a}i\gamma_5u^a({\bf P'})$, 
 ${\overline u}^d({\bf P})(p_N)_d i\gamma_5u({\bf p_N})$,
${\overline u}^b({\bf P'})g_{bc}(P+P')^{\mu}u^c({\bf P})$,
 ${\overline u}^b({\bf P'})g_{bc}\Bigl[-i\sigma^{\mu\nu}k'_{\nu}
\Bigr]u^c({\bf P})$ and 
${\overline u}^b({\bf P'})g_{bc}u^c({\bf P})$
which are needed for 
the $\Delta-\gamma\Delta$-type vertex 
in (3.6) depicted in Fig. 2B.

\vspace{0.05cm}

\begin{center}

{\em C.  The double $\Delta$ exchange amplitude
and transitions  
$({\cal E_L}^{3/2})_{\gamma'\pi'  N'-\pi N}^{\mu}(\Delta-pole)\Longrightarrow$
$({\cal E_L}^{3/2})_{\gamma'\pi'  N'-\pi N}^{\mu}(\Delta\Delta)$ and 
${\cal B}^{3/2}_{\pi'  N'-\pi N}(\Delta-pole)\Longrightarrow$
${\cal B}^{3/2}_{\pi'  N'-\pi N}(\Delta\Delta)$}.

\end{center}

\vspace{0.15cm}

Next we have to extract from 
${\cal E_L}^{3/2})_{\pi' N'-\pi N}^{\mu}({\Delta}-pole)$
(A10b) the amplitude which has the same analytical properties
as the double $\Delta$ exchange term in Fig. 2B. 
Using a simple algebra we rewrite  (A.11b) as

$${\cal V}_H=
{1\over{
\Bigl({p'}_{\pi}^{o}+{p'}_{N}^{o}-{P'}^o_{\Delta}(s')\Bigr)
\Bigl({p}_{\pi}^{o}+{p}_{N}^{o}-{P}^o_{\Delta}(s)\Bigr) }}
\Biggl\{ R_+\biggl[
{{|{\bf k'}|}\over {s-s'}}
-{ {{P}^o_{\Delta}(s)-{P'}^o_{\Delta}(s')}\over{s-s'}}
\biggr]
$$
$$+{{R_-}\over {s-s'} }
\biggl[ 
{p}_{\pi}^{o}+{p_{N}}^{o}+{p'}_{\pi}^{o}+{p'_{N}}^{o}
-{P}^o_{\Delta}(s)-{P'}^o_{\Delta}(s')
\biggr]\Biggr\},
\eqno(A.12a)$$

where 
$$R_{\pm}
={1\over 2}\Bigl[\mu_{N'}{\cal R}_{N'}
\pm\mu_{N}{\cal R}_{N}\Bigr]\eqno(A.12b)$$
and we use the identities
$1/a\pm1/b=1/a\ (b\pm a)\ 1/b$ with
$a= \Bigl({p'}_{\pi}^{o}+{p'}_{N}^{o}-{P'}^o_{\Delta}(s')\Bigr)$
and
$b= \Bigl({p}_{\pi}^{o}+{p}_{N}^{o}-{P}^o_{\Delta}(s)\Bigr)$.
These transformations play a central role in
 connection between the amplitudes in Fig. 1 and in Fig. 2B.

The first part in (A.12a) is regular at $|{\bf k'}|=0$, where
$s=s'$, because $\Bigl({P}^o_{\Delta}(s)-{P'}^o_{\Delta}(s')\Bigr)/(s-s')$
is  finite at $s=s'$. 

The second part of ${\cal V}_H$ (A.12a) can describe only one 
$\Delta$ exchange because
${p}_{\pi}^{o}+{p_{N}}^{o}+{p'}_{\pi}^{o}+{p'_{N}}^{o}
-{P}^o_{\Delta}(s)-{P'}^o_{\Delta}(s')$ cancels out one of the $\Delta$
propagators. This expression 
has different behavior in two different cases:

${\bf 1.}$  For charge exchange reactions $R_-/ (s-s')$ can be singular at 
the threshold $|{\bf k'}|=0$. This case needs a special investigation
which is out of the scope of this article.

${\bf 2.}$ For the  $\pi N$ bremsstrahlung reactions without charge exchange
(e.g. $\pi^{\pm} p\longrightarrow\gamma \pi^{\pm} p$ or 
$\pi^{o} p\longrightarrow\gamma \pi^{o} p$) 
we have $e_{N'}=e_N$ and 
 $\mu_{N'}=\mu_N$. In this case     
$R_-/ (s-s')$ is finite at the threshold $|{\bf k'}|=0$ and this part
corresponds to
the one $\Delta$ exchange diagrams of the $\pi N\to \gamma'\pi' N'$
reaction.
Using identity (A.1) one can  separate one and double $\Delta$ exchange
contributions through 
the different  current conservation conditions, i.e., one can  split 
$({\cal E_L}^{3/2})_{\gamma'\pi' N'-\pi N}^{\mu}({\Delta}-pole)$  and 
${\cal B}^{3/2}_{\pi' N'-\pi N}({\Delta}-pole)$  into two parts

$$
({\cal E_L}^{3/2})_{\gamma'\pi' N'-\pi N}^{\mu}({\Delta}-pole)=
({\cal E_L}^{3/2})_{\gamma'\pi' N'-\pi N}^{\mu}(\Delta\Delta)+
({\cal E_L}^{3/2})_{\gamma'\pi' N'-\pi N}^{\mu}(\Delta)$$
$${\cal B}^{3/2}_{\pi' N'-\pi N}({\Delta}-pole)=
{\cal B}^{3/2}_{\pi' N'-\pi N}(\Delta\Delta)+
{\cal B}^{3/2}_{\pi' N'-\pi N}(\Delta),\eqno(A.13)$$

where

$$({\cal E_L}^{3/2})_{\gamma'\pi' N'-\pi N}^{\mu}({\Delta}\Delta)=
 {1\over{ |{\bf p'}_N||{\bf p}_N| }}
{\overline u}({\bf p'_N})(p'_N)_{a}i\gamma_5u^a({\bf P'})$$
$$\Biggl\{
{\overline u}^b({\bf P'})g_{bc}
\Bigl[(P+P')^{\mu}{\cal V}_E^{(+)}-i\sigma^{\mu\nu}k'_{\nu}{\cal V}_H^{(+)}
\Bigr]
u^c({\bf P})\Biggr\}
{\overline u}^d({\bf P})(p_N)_d i\gamma_5u({\bf p_N}),\eqno(A.14a)$$

$$({\cal E_L}^{3/2})_{\gamma'\pi' N'-\pi N}^{\mu}({\Delta})=
 {1\over{ |{\bf p'}_N||{\bf p}_N| }}
{\overline u}({\bf p'_N})(p'_N)_{a}i\gamma_5u^a({\bf P'})$$
$$\Biggl\{
{\overline u}^b({\bf P'})g_{bc}\Bigl[
(P+P')^{\mu}{\cal V}_E^{(-)}-i\sigma^{\mu\nu}k'_{\nu}{\cal V}_H^{(-)}
\Bigr]u^c({\bf P})\Biggr\}
{\overline u}^d({\bf P})(p_N)_d i\gamma_5u({\bf p_N}),\eqno(A.14b)$$

$${\cal B}^{3/2}_{\pi' N'-\pi N}({\Delta}\Delta)=
 {{s-s'}\over{ |{\bf p'}_N||{\bf p}_N| }}
{\overline u}({\bf p'_N})(p'_N)_{a}i\gamma_5u^a({\bf P'})
\Biggl\{
{\overline u}^b({\bf P'})g_{bc}
{\cal V}_E^{(+)}u^c({\bf P})\Biggr\}
{\overline u}^d({\bf P})(p_N)_d i\gamma_5u({\bf p_N}),\eqno(A.14c)$$

$${\cal B}^{3/2}_{\pi' N'-\pi N}({\Delta})=
 {{s-s'}\over{ |{\bf p'}_N||{\bf p}_N| }}
{\overline u}({\bf p'_N})(p'_N)_{a}i\gamma_5u^a({\bf P'})
\Biggl\{
{\overline u}^b({\bf P'})g_{bc}
{\cal V}_E^{(-)}u^c({\bf P})\Biggr\}
{\overline u}^d({\bf P})(p_N)_d i\gamma_5u({\bf p_N}),\eqno(A.14d)$$

where

$${\cal V}_E^{(+)}={{
e_N({\cal R}_{N'}+{\cal R}_{N})+e_{\pi}({\cal R}_{\pi'}+{\cal R}_{\pi})
}\over{2(s-s')}}
{{
|{\bf k'}|- ({P}^o_{\Delta}(s)-{P'}^o_{\Delta}(s'))
}\over{
\Bigl({p'}_{\pi}^{o}+{p'}_{N}^{o}-{P'}^o_{\Delta}(s')\Bigr)
\Bigl({p}_{\pi}^{o}+{p}_{N}^{o}-{P}^o_{\Delta}(s)\Bigr) }}\eqno(A.15a)$$

$${\cal V}_H^{(+)}={{
\mu_N({\cal R}_{N'}+{\cal R}_{N})}\over{2(s-s')}}
{{|{\bf k'}|
-({P}^o_{\Delta}(s)-{P'}^o_{\Delta}(s') )}\over{
\Bigl({p'}_{\pi}^{o}+{p'}_{N}^{o}-{P'}^o_{\Delta}(s')\Bigr)
\Bigl({p}_{\pi}^{o}+{p}_{N}^{o}-{P}^o_{\Delta}(s)\Bigr) }}\eqno(A.15b)$$

$${\cal V}_E^{(-)}={{
e_N({\cal R}_{N'}-{\cal R}_{N})+e_{\pi}({\cal R}_{\pi'}-{\cal R}_{\pi})
}\over{2(s-s')}}
{{
{p'}_{\pi}^{o}+{p'}_{N}^{o}-{P'}^o_{\Delta}(s')
+{p}_{\pi}^{o}+{p}_{N}^{o}-{P}^o_{\Delta}(s)}\over{
\Bigl({p'}_{\pi}^{o}+{p'}_{N}^{o}-{P'}^o_{\Delta}(s')\Bigr)
\Bigl({p}_{\pi}^{o}+{p}_{N}^{o}-{P}^o_{\Delta}(s)\Bigr) }}\eqno(A.15c)$$

$${\cal V}_H^{(-)}={{
\mu_N({\cal R}_{N'}-{\cal R}_{N})}\over{2(s-s')}}
{{
{p'}_{\pi}^{o}+{p'}_{N}^{o}-{P'}^o_{\Delta}(s')
+{p}_{\pi}^{o}+{p}_{N}^{o}-{P}^o_{\Delta}(s)}\over{
\Bigl({p'}_{\pi}^{o}+{p'}_{N}^{o}-{P'}^o_{\Delta}(s')\Bigr)
\Bigl({p}_{\pi}^{o}+{p}_{N}^{o}-{P}^o_{\Delta}(s)\Bigr) }}\eqno(A.15d)$$

With (A.1) it easy to check that

$$ 
{k'}_{\mu}({\cal E_L}^{3/2})_{\gamma'\pi' N'-\pi N}^{\mu} ({\Delta}{\Delta})=-
{\cal B}^{3/2}_{\pi' N'-\pi N}({\Delta}{\Delta});\ \ \
{k'}_{\mu}({\cal E_L}^{3/2})_{\gamma'\pi' N'-\pi N}^{\mu} ({\Delta})=-
{\cal B}^{3/2}_{\pi' N'-\pi N}({\Delta}).
\eqno(A.16)$$
 
Thus we have extracted the double $\Delta$ exchange
part from the $\Delta-pole$  amplitudes (A.10b,c).
Afterwards,  the Ward-Takahashi identity (A.10a)  is divided into 
two independent identities (A.16).
The resulting Ward-Takahashi identity   (A.16) 
contains the double $\Delta$ exchange terms.

\vspace{0.05cm}

\begin{center}

{\em D.  
An alternative form of the double $\Delta$  exchange amplitude 
 $({\cal E_L}^{3/2})_{\gamma'\pi'  N'-\pi N}^{\mu}(\Delta\Delta)$}.

\end{center}
\vspace{0.15cm}

 Hereafter it is convenient to represent
$({\cal E_L}^{3/2})_{\gamma'\pi'  N'-\pi N}^{\mu}(\Delta\Delta)$ 
(A.14a) and ${\cal B}^{3/2}_{\pi' N'-\pi N}({\Delta}\Delta)$ (A.14c)
through the  
$\pi N\to \Delta$, $\Delta\to\gamma'\Delta'$, $\Delta'\to\pi' N'$
vertices (3.5a,b)
and the intermediate $\Delta$ propagators. Therefore 
we rewrite (A.15a,b) as

$${\cal V}^{(+)}_E={1\over{
\Bigl({p'}_{\pi}^{o}+{p'}_{N}^{o}-{P'}^o_{\Delta}(s')\Bigr)
\Bigl({p}_{\pi}^{o}+{p}_{N}^{o}-{P}^o_{\Delta}(s)\Bigr) }}
{\rm g}_{\pi' N'-\Delta'}(s',k'){\sc V}_E
{\rm g}_{\Delta-\pi N}(s),
\eqno(A.17a)$$

$${\cal V}^{(+)}_H={1\over{
\Bigl({p'}_{\pi}^{o}+{p'}_{N}^{o}-{P'}^o_{\Delta}(s')\Bigr)
\Bigl({p}_{\pi}^{o}+{p}_{N}^{o}-{P}^o_{\Delta}(s)\Bigr) }}
{\rm g}_{\pi' N'-\Delta'}(s',k'){\sc V}_H
{\rm g}_{\Delta-\pi N}(s).
\eqno(A.17b)$$

Relations (A.17a,b) allows to rewrite (A.14a,c)  as

$$
({\cal E_L}^{3/2})_{\gamma'\pi' N'-\pi N}^{\mu}(\Delta\Delta)=
{{<{\bf p'}_N,{\bf p'}_{\pi}|{\sl g}_{\pi' N'-\Delta'}|{\bf P'}_{\Delta}>}
\over{{p'}_{\pi}^{o}+{p'}_{N}^{o}-{P'}^o_{\Delta}(s')}}$$
$$\Biggl\{
{\overline u}^b({\bf P'}_{\Delta})g_{bc}\Bigl[
(P+P')^{\mu}{\sc V}_E-i\sigma^{\mu\nu}k'_{\nu}{\sc V}_H
\Bigr]u^c({\bf P}_{\Delta})\Biggr\}
{{<{\bf P}_{\Delta}|{\sl g}_{\Delta-\pi N}|{\bf p}_N,{\bf p}_{\pi}>}
\over{{p}_{\pi}^{o}+{p}_{N}^{o}-{P}^o_{\Delta}(s) }},
\eqno(A.18a)$$

$$
{\cal B}^{3/2}_{\pi' N'-\pi N}(\Delta\Delta)=-
{{<{\bf p'}_N,{\bf p'}_{\pi}|{\sl g}_{\pi' N'-\Delta'}|{\bf P'}_{\Delta}>}
\over{{p'}_{\pi}^{o}+{p'}_{N}^{o}-{P'}^o_{\Delta}(s')}}$$
$$\Biggl\{
{\overline u}^b({\bf P'}_{\Delta})g_{bc}\Bigl[
(s-s'){\sc V}_E
\Bigr]u^c({\bf P}_{\Delta})\Biggr\}
{{<{\bf P}_{\Delta}|{\sl g}_{\Delta-\pi N}|{\bf p}_N,{\bf p}_{\pi}>}
\over{{p}_{\pi}^{o}+{p}_{N}^{o}-{P}^o_{\Delta}(s) }}.
\eqno(A.18b)$$

The amplitude (A.18a) has the 
form of the usual $\Delta$ radiation diagram in Fig. 2B with
the $\Delta-\gamma\Delta$ vertex function
$(P+P')^{\mu}{\sc V}_E-i\sigma^{\mu\nu}k'_{\nu}{\sc V}_H$
 instead of the $\Delta-\gamma \Delta$ 
vertex (B.3a,b) in Appendix B.


\vspace{0.05cm}

\begin{center}

{\em E. Transitions 
$({\cal E_L}^{3/2})_{\gamma'\pi'  N'-\pi N}^{\mu}(\Delta\Delta)\Longrightarrow$
$({\cal E_L}^{3/2})_{\gamma'\pi'  N'-\pi N}^{\mu}(\Delta-\gamma\Delta)$
 and
${\cal B}^{3/2}_{\pi'  N'-\pi N}(\Delta\Delta)\Longrightarrow$
${\cal B}^{3/2}_{\pi'  N'-\pi N}(\Delta-\gamma\Delta)$}

\end{center}
\vspace{0.15cm}

It is important to note that 
${\cal I}_{\gamma'\pi' N'-\pi N}^{\mu}(\Delta-\gamma\Delta)$ (3.6) and
$({\cal E_L}^{3/2})_{\gamma'\pi' N'-\pi N}^{\mu}(\Delta\Delta)$ (A.18a)
contain different $\Delta-\gamma\Delta$ vertices.
In order to unify these vertex functions
we extract from  
 $({\cal E_L}^{3/2})_{\gamma'\pi' N'-\pi N}^{\mu}(\Delta\Delta)$ (3.15d)
the part with the $\Delta-\gamma\Delta$ vertex from (3.6)

$$({\cal E_L}^{3/2})_{\gamma'\pi' N'-\pi N}^{\mu}({\Delta}\Delta)
=({\cal E_L}^{3/2})_{\gamma'\pi' N'-\pi N}^{\mu}({\Delta}-\gamma\Delta)
+({ E_L}^{3/2})_{\gamma'\pi' N'-\pi N}^{\mu}(\Delta),
\eqno(A.19a)$$

$$({\cal B}^{3/2})_{\pi' N'-\pi N}^{\mu}({\Delta}\Delta)
=({\cal B}^{3/2})_{\pi' N'-\pi N}^{\mu}({\Delta}-\gamma\Delta)
+({ B}^{3/2})_{\pi' N'-\pi N}^{\mu}(\Delta).
\eqno(A.19b)$$

where $({\cal E_L}^{3/2})_{\gamma'\pi' N'-\pi
  N}^{\mu}({\Delta}-\gamma\Delta)$ and
$({\cal B}^{3/2})_{\gamma'\pi' N'-\pi N}^{\mu}({\Delta}-\gamma\Delta)$
are defined by (3.3a,b) and

$$({ E_L}^{3/2})_{\gamma'\pi' N'-\pi N}^{\mu}(\Delta)=
{{<{\bf p'}_N,{\bf p'}_{\pi}|{\sl g}_{\pi' N'-\Delta'}|{\bf P'}_{\Delta}>}
\over{{p'}_{\pi}^{o}+{p'}_{N}^{o}-{P'}^o_{\Delta}(s')}}$$
$$\Biggl\{
{\overline u}^b({\bf P'}_{\Delta})g_{bc}\Bigl[
(P+P'-P_{\Delta}-P'_{\Delta})^{\mu}{\sc V}_E-\
i\sigma^{\mu\nu}({k'-k'_{\Delta}})_{\nu}{\sc V}_H
\Bigr]u^c({\bf P}_{\Delta})
{{<{\bf P}_{\Delta}|{\sl g}_{\Delta-\pi N}|{\bf p}_N,{\bf p}_{\pi}>}
\over{{p}_{\pi}^{o}+{p}_{N}^{o}-{P}^o_{\Delta}(s) }}.\eqno(A.20a)$$

$${ B}^{3/2}_{\pi' N'-\pi N}(\Delta)=
{{<{\bf p'}_N,{\bf p'}_{\pi}|{\sl g}_{\pi' N'-\Delta'}|{\bf P'}_{\Delta}>}
\over{{p'}_{\pi}^{o}+{p'}_{N}^{o}-{P'}^o_{\Delta}(s')}}$$
$$ \Biggl\{
{\overline u}^b({\bf P'}_{\Delta})g_{bc}\Bigl[
k'_o(P+P'-P_{\Delta}-P'_{\Delta})^{o}{\sc V}_E-
ik'_{\mu}\sigma^{\mu o}(k'-k'_{\Delta})_{o}{\sc V}_H
\Bigr]u^c({\bf P}_{\Delta})\Biggr\}
{{<{\bf P}_{\Delta}|{\sl g}_{\Delta-\pi N}|{\bf p}_N,{\bf p}_{\pi}>}
\over{{p}_{\pi}^{o}+{p}_{N}^{o}-{P}^o_{\Delta}(s) }}
.\eqno(A.20b)$$

 The difference between 
$ ({\cal E_L}^{3/2})_{\gamma'\pi'N'-\pi N}^{\mu}(\Delta\Delta)$ (A.18a) 
and $({\cal E_L}^{3/2})_{\gamma'\pi' N'-\pi N}^{\mu}({\Delta}-\gamma\Delta)$ 
(3.3a) 
makes the zero components of the kinematic factors
$(P+P')^{\mu}-(P_{\Delta}+P'_{\Delta})^{\mu}=\delta^{\mu0}
\Bigl[P^o+{P'}^{o}-P_{\Delta}^o-{P'}^o_{\Delta}\Bigr]$;
$(k'-k'_{\Delta})_{\nu}=\delta_{\nu 0}
\Bigl[P^o-{P'}^{o}-P_{\Delta}^o+{P'}^o_{\Delta}\Bigr]$. 
These kinematic factors 
cancel out one of the $\Delta$ propagators 
$1/({P'}^{o}-{P'}^o_{\Delta})$ or $1/({P}^{o}-{P}^o_{\Delta})$
in $({{ E_L}}^{3/2})_{\gamma'\pi' N'-\pi N}^{\mu}(\Delta)$ (3.20c).
Therefore $({{ E_L}}^{3/2})_{\gamma'\pi' N'-\pi N}^{\mu}(\Delta)$ (A.20b)
corresponds to the one $\Delta$ exchange term.

Modification of  
$({\cal E_L}^{3/2})_{\gamma'\pi'N'-\pi  N}^{\mu}(\Delta\Delta)$ (3.3a)
and ${\cal B}^{3/2}_{\pi'N'-\pi N}(\Delta\Delta)$ (3.3b)
according to  (3.19a,b) generates two new  
Ward-Takahashi identities: identity (3.2) 
for the amplitudes  (3.3a,b) and  
$k'_{\mu}({ E_L}^{3/2})_{\gamma'\pi' N'-\pi N}^{\mu}(\Delta)
=-{ B}^{3/2}_{\pi' N'-\pi N}(\Delta)$
for
the amplitudes (A.20a,b).

An additional expression
like $({ E_L}^{3/2})_{\gamma'\pi' N'-\pi N}^{\mu}(\Delta)$ 
(A.20a) was also used  in other papers \cite{Mink,Heller}
in order to ensure  current
conservation for the total $\pi N$ bremsstrahlung amplitude.
Unlike  the case in those papers,  
$({ E_L}^{3/2})_{\gamma'\pi' N'-\pi N}^{\mu}(\Delta)$  corresponds to
the one-$\Delta$ exchange amplitude which satisfies the 
independent Ward-Takahashi identity.


\vspace{0.05cm}

\begin{center}
   {{\bf Appendix B:} $\gamma'\Delta'-\Delta$ vertex function with on mass 
shell $\Delta$'s}

\end{center}

\vspace{0.15cm}

The  $\Delta-\gamma\Delta$ vertices  can be  
constructed using the analytical continuation 
of  the spin $3/2$ particle electromagnetic vertex function
$<out;{\bf P'}|{\cal J}^{\mu}(0)|{\bf P};in>$ in the complex region.
The electromagnetic vertex of the spin $3\over 2$ particles with the
real mass $m_{3/2}$ is

$$<out;{\bf P'}|{\cal J}^{\mu}(0)|{\bf P};in>=(P+P')^{\mu}\Biggl(
{\overline u}^{\sigma}({\bf P'})
\Bigl[g_{\rho\sigma}G_1({k'}^2)
+{ { {k'}_{\sigma}{k' }_{\rho}}\over{m_{3\over 2}^2}}G_3({k'}^2)
\Bigr]\Biggr)u^{\rho}({\bf P})$$
$$+{\overline u}^{\sigma}({\bf P'})\Biggl(
-i\sigma^{\mu\nu}{k'}_{\nu}\Bigl[g_{\rho\sigma}G_2({k'}^2)
+{ { {k'}_{\sigma}{k'}_{\rho}}\over{m_{3\over 2}^2}}G_4({k'}^2)
\Bigr]\Biggr)u^{\rho}({\bf P}),\eqno(B.1)$$

where ${k'}_{\mu}=(P-P')_{\mu}$ denotes the four momentum of the 
emitted photon,
${\bf P}={\bf p}_N+{\bf p}_{\pi}=
{\bf P}_{\Delta}$, ${P}^o=\sqrt{m_{3\over 2}^2+{\bf P}^2}$;\ \ \ 
 ${\bf P'}={\bf p}'_N+{\bf p'}_{\pi}={\bf P'}_{\Delta}$, 
 ${P'}^o=\sqrt{m_{3\over 2}^2+{\bf P'}^2}$ are the four moments of 
 spin $3/2$
particles with a mass $m_{3\over 2}$ in the initial and final states.

Expression (B.1) can be analytically continued  for the 
unequal masses of the 
particles in the ``in'' ($m_{3/2}(in)=m_{3/ 2}$) and in the "out'' 
($m_{3/2}(out)=m'_{3/2}$) states

$$<out;{\bf P'} m'_{3\over 2}|{\cal J}^{\mu}(0)|{\bf P}m_{3\over
  2};in>
=(P+P')^{\mu}\Biggl(
{\overline u}^{\sigma}({\bf P'})
\Bigl[g_{\rho\sigma}G_1({k'}^2,m^2_{3\over 2},m'^2_{3\over 2})
+{ { {k'}_{\sigma}{k' }_{\rho}}\over
  {m'_{3\over 2}m_{3\over 2}}}G_3({k'}^2,m^2_{3\over 2}
,m'^2_{3\over 2})\Bigr]\Biggr)u^{\rho}({\bf P})$$
$$+{\overline u}^{\sigma}({\bf P'})\Biggl(
-i\sigma^{\mu\nu}{k'}_{\nu}\Bigl[g_{\rho\sigma}G_2({k'}^2,
m^2_{3\over 2},m'^2_{3\over 2})
+{ { {k'}_{\sigma}{k' }_{\rho}}\over
  {m'_{3\over 2}m_{3\over 2}}}G_4({k'}^2,m^2_{3\over 2}
,m'^2_{3\over 2})
\Bigr]\Biggr)u^{\rho}({\bf P}),\eqno(B.2)$$
where for $m'_{3/2}=m_{3/2}$ expression (B.2) coincides with (B.1).

The extension of (B.2)
in the complex energy and mass region of the $\Delta$ resonance  implies
the replacements
$m_{3/2}\Longrightarrow {\sf m}_{\Delta}(s)$ and
$m'_{3/2}\Longrightarrow {\sf m}_{\Delta}(s')$,
where ${\sf m}_{\Delta}$ is given in  (3.9b).
Correspondingly, we obtain the complex energies (3.9c,d)
and the complex zero component of the four-vector
of the four momentum transfer ${\bf k'}={\bf k'}_{\Delta}$, 
$k'_o\Longrightarrow (k'_{\Delta})_o=
\sqrt{{\sf m}^2(s)+{\bf P}^2_{\Delta}}
-\sqrt{{\sf m}^2(s')+{\bf P'}^2_{\Delta}}$.

The general double $\Delta$ exchange term 
${\cal I}_{\gamma'\pi' N'-\pi  N}^{\mu}({\Delta}-\gamma{\Delta})$ (3.18)
contains the following full $\Delta -\gamma'\Delta'$ 
vertex function $<{\bf P'}_{\Delta},s'|J_{\mu}(0)|{\bf
  P}_{\Delta},s>$ with on mass shell $\Delta$'s

\footnotetext{ 
Another double $\Delta$ exchange term
contains the $\Delta-\pi'\Delta'$ vertex function.
But this term in negligible small\cite{Mink}. }

$$
<{\bf P'}_{\Delta},{\sf m}_{\Delta}(s')|{\cal J}^{\mu}(0)
|{\bf P}_{\Delta},{\sf m}_{\Delta}(s)>
=(P_{\Delta}+P'_{\Delta})^{\mu}\Biggl(
{\overline u}^{\sigma}({\bf P'}_{\Delta})
\Bigl[g_{\rho\sigma}
G_1({k'}_{\Delta}^2,s,s')$$
$$+{ {{k'_{\Delta}}_{\sigma}{k'_{\Delta} }_{\rho}}\over{M^2_{\Delta}}}
G_3({k'}_{\Delta}^2,s,s')
\Bigr]u^{\rho}({\bf P}_{\Delta})\Biggr)$$
$$+{\overline u}^{\sigma}({\bf P'}_{\Delta})\Biggl(
-i\sigma^{\mu\nu}{k'_{\Delta}}_{\nu}\Bigl[g_{\rho\sigma}
G_2({k'}_{\Delta}^2,s,s')
+{ {{k'_{\Delta}}_{\sigma}{k'_{\Delta} }_{\rho}}\over{M^2_{\Delta}}}
G_4({k'}_{\Delta}^2,s,s')
\Bigr]\Biggr)u^{\rho}({\bf P}_{\Delta}),\eqno(B.3)$$

where we introduced the auxiliary four-vector
${k'_{\Delta}}_{\mu}=(P_{\Delta}-P'_{\Delta})_{\mu}$ for the momentum transfer
 and $g_{\mu\nu}$ is the metric tensor. An additional dependence of
the form factors  $G_i({k'}_{\Delta}^2,s,s')$ on the 
variables $s$ and $s'$ 
is generated by  
${\sf m}_{\Delta}(s)$ and ${\sf m}_{\Delta}(s')$ (3.4d).

The form factors 
$G_i({k'}_{\Delta}^2,s,s')$ 
are simply connected with the charge monopole
$G_{C0}$, magnetic dipole $G_{M1}$, 
electric quadrupole
$G_{E2}$ and magnetic octupole
$G_{M3}$ form factors of the $\Delta$ resonance.
{\footnotemark}

The terms 
$\sim {k'}_{\Delta}^2/4M_{\Delta}^2$ in the 
$\Delta-\gamma'\Delta'$ vertex for the
low energy photons can be neglected
and only terms $\sim 1/M_{\Delta}$ can be taken into account. 
Then (B.3) reduces to (3.7).

\footnotetext{ 
Other choices of $G_i$ form factors are considered in ref. \cite{PasVan3} }

An important property of the electromagnetic $\Delta$
vertices (B.3) and (3.7) is that at the threshold ($k'=0$ and $k'_{\Delta}=0$)
they coincide with  $G_i(0,m^2_{3\over 2},m'^2_{3\over 2})$ in (B.2). 
But the form factors $G_i({k'}^2,m^2_{3\over 2},m'^2_{3\over 2})$ in (B.2)
 are real according to the $C,P,T$ invariance. 
Consequently, the form factors  $G_i({k'}_{\Delta}^2,s,s')$ 
at the threshold are also real. 
Therefore,   $G_{C0}({k'}_{\Delta}^2,s,s')$ and 
$G_{M1}({k'}_{\Delta}^2,s,s')$ satisfy the
following normalization conditions

$$G_{C0}({k'}_{\Delta}^2=0,s,s)=e_{\Delta};\ \ \
 G_{M1}({k'}_{\Delta}^2=0,s,s)=\mu_{\Delta},\eqno(B.4a)$$
where $e_{\Delta}$ and $\mu_{\Delta}$ denote the charge and 
magnetic dipole moment of the $\Delta$'s. 

The exact form of vertex functions (B.3a,b) ensure the validity 
of the special one-body current conservation condition  
for  the $\Delta$ vertex function

$${k'}_{\Delta}^{\mu}<{\bf P'}_{\Delta},{\sf m}_{\Delta}
(s')|J_{\mu}(0)|{\bf
  P}_{\Delta},{\sf m}_{\Delta}(s)>=$$
$$ \left \{ \begin{array}{llllllllllllll} 0\ \ \ \ \ \ \ \ \ 
\ \ \ \ \ \ \ \ \ \ \ \ \ \ \ \ \ \ \ \ \ \ \ \ \ \ \ \ \ \ \
\ \ \ \ \ \ \ \ \ \ \ \ \ \ \ \ \ \ \ for\ constant \ 
{\sf  m}_{\Delta}\ (3.4c)   \\
{{ {\sf m}_{\Delta}^2(s)-
{\sf m}_{\Delta}^2(s') }
\over {2 M_{\Delta}}}
G_{C0}({k'}_{\Delta}^2,s,s') {\overline u}^{\sigma}({\bf P'}_{\Delta})
g_{\rho\sigma}u^{\rho}({\bf P}_{\Delta}) \ \ \  for\ {\sf m}_{\Delta}(s)\ 
(3.4d)
\end{array}\right.\eqno(B.4b)$$

Equation (B.4b) expresses  the analytical continuation of usual 
current conservation  for the real spin $3/2$ particle
vertex function in the complex energy-mass region of the $\Delta$'s.
Certainly, this ``one-body intermediate
$\Delta$ current'' conservation is not necessary for real 
current conservation of the full $\pi N$ radiation amplitude.    

It must be emphasized that the present formulation
of the  $\Delta$ degrees of freedom 
does not use a Heisenberg local field operator
of the $\Delta$ resonance or a Lagrangian with the local 
$\Delta$ field operators. This simplifies the formulation because 
 it is not possible to construct a Fock space for a ``free''
resonance state with a complex mass and a complex energy.
A renormalization procedure for
 real spin $3/2$ states can generate  intermediate $\Delta$
complex states. This renormalization is equivalent to the extension
of the vertex functions (B.1) or (B.2) into the complex $\Delta$
vertex (B.3). 
In the present approach we use the vertices only with the on mass shell 
$\Delta$'s. Therefore,  ambiguities generated 
by unphysical gauge transformations of the $\Delta$-particle field operator
$\Psi_{\Delta}^a\longrightarrow \Psi_{\Delta}^a+C\gamma^a\gamma_b 
\Psi_{\Delta}^b$ \cite{Bammer} with an arbitrary parameter $C$
do not appear in the present formulation.
Sensitivity of the $\gamma p\to 
\gamma'\pi' p'$ observable to the choice of the form of the intermediate 
$\Delta$ propagator is demonstrated in \cite{MF}.

In the off  mass shell region, where  ${P'}^2\ne m^2_{\Delta}$ and 
${P}^2\ne m^2_{\Delta}$,
the $\Delta-\gamma'\Delta'$ vertex is a function 
of  two independent four moments of each $\Delta$.
 Therefore, for the off mass shell $\Delta$'s
(B.3) and (3.7) take a much more complicated 
forms with the increasing number of the form factors $G_i$, because 
each of the conditions
$P^2_{\Delta}\ne m_{\Delta}^2$ and  
$(i\gamma_{\sigma}P_{\Delta}^{\sigma}\ne m_{\Delta}^2)$ 
reduplicates  the  number of the form factors. Therefore, instead 
of two form factors in (3.7) we get $8$ form factors for the off mass shell
$\Delta-\gamma'\Delta'$  vertices. The role of these six 
additional form factors is as important as 
the contribution of  the off shell effects like the mass and charge
renormalization. 
In addition, these form factors of the $\Delta-\gamma'\Delta'$ vertex
with off mass shell $\Delta$'s depend on three complex variables 
${k'}^2_{\Delta}$, $P^2_{\Delta}$ and  ${P'}^2_{\Delta}$.
Therefore, the use of the off 
mass shell $\Delta$ propagators together with the on mass shell 
$\Delta-\gamma'\Delta'$, as is done in refs. \cite{Amiri,PasVan2,PasVan3},
is inconsistent.


\vspace{0.05cm}

\begin{center}
   {{\bf Appendix C:} On mass shell $\Delta$ degrees of freedom and
     construction of the double $\Delta$ exchange term 
in Fig. 2B.}

\end{center}

\vspace{0.15cm}

The on mass shell intermediate $\Delta$ states are usually introduced 
via the  $\Delta$ resonance pole position
in the $\pi N$ amplitude.
We shall shortly consider  the corresponding formulation
within the time-ordered field-theoretical 
approach \cite{Ann,MF,NP}. In this formulation
the off mass shell $\pi N$ amplitudes (2.9a,b,c,d)
 are simply connected with the $\pi N$ $t$-matrix 
${\sc T}({\bf p'_Np'_{\pi},p_Np_{\pi}};E)$, 
 which satisfies the relativistic Lippmann-Schwinger-type
 equation in the c.m. frame 

$${\sc T}({\bf p',p};E_{\bf p})={\sc U}({\bf p',p};E_{\bf p})-
\int  {\sc U}({\bf p',q};E_{\bf p}){ {d^3{\bf q}}\over
{E_{\bf p}-E_{\bf q}-i\epsilon}}
{\sc T}({\bf q,p};E_{\bf p}),\eqno(C.1)$$

where
$E_{\bf p}\equiv P_o=\sqrt{ {\bf p}^2+m_{\pi}^2}+\sqrt{ {\bf p}^2+m_{N}^2}$
and ${\bf p}$  are the $\pi N$ energy and the relative 
three-momentum in c.m. frame. 
Equation (C.1) can be symbolically represented as  

$${\sc T}(E_{\bf p})={\sc U}(E_{\bf p})+
{\sc U}(E_{\bf p}){\sc G_o}(E_{\bf p}){\sc T}(E_{\bf p})
={\sc U}(E_{\bf p})+
{\sc U}(E_{\bf p}){\cal G}_{\pi N}(E_{\bf p}){\sc U}(E_{\bf p}),
\eqno(C.2)$$

where ${\sc G_o}(E_{\bf p})$ and ${\cal G}_{\pi N}(E_{\bf p})$ are the
  free and total Green functions of the $\pi N$ system
and ${\sc U}(E)\equiv {\sc U}({\bf p',p};E)=
{\sc A}({\bf p',p})+E{\sc B}({\bf p',p})$ is the linear energy
depending on the field-theoretical potential with
a Hermitian ${\sc A}({\bf p',p})$ and ${\sc B}({\bf p',p})$ matrices.
The full $\pi N$ Green function satisfies the completeness condition

$${\cal G}_{\pi N }(E)=\sum_{\pi N} 
{ {|{ \Psi}_{\pi N}({\bf q})>
d^3{\bf q}<\Psi_{\pi N}({\bf q})|(1-{\sc B}) }\over{E-E_{\bf q}-i\epsilon}},
\eqno(C.3)$$

where ${ \Psi}_{\pi N}({\bf q})$ denotes the $\pi N$ wave function
which can be determined via the solution of  (C.1).

The $\Delta$ resonance pole in the complex energy region generates
the following representation of the full $\pi N$ wave function

$${\cal G}_{\pi N }(E)=\sum_{\Delta}
{ {| \Psi_{\Delta}><\Psi_{\Delta}|(1-{\sc B}) }\over{E-E_{\Delta}}}
+\ non-pole\ part,\eqno(C.4a)$$

where $m_{\Delta}=1232MeV - {i\over  2}120MeV$ and
$E_{\Delta}\equiv P^o_{\Delta}=\sqrt{m_{\Delta}^2+{\bf P}^2}$
according to (3.9a) and (3.9c).
$m_{\Delta}$ indicates the $\Delta$-resonance 
pole position of the full $\pi N$ Green function or the total $\pi N$
amplitude. 

Using (C.4) and (C.2) one can extract 
the $\Delta$ exchange part  of the $\pi N$ $t$-matrix   

$$\biggl[{\sc T}(E)\biggr]_{one\ \Delta\ exchange\ part}
=\sum_{\Delta}
{ {{\sc U}(E) |\Psi_{\Delta}><\Psi_{\Delta}|(1-{\sc B})
{\sc U}(E) }\over{E-E_{\Delta}}}\eqno(C.4b)$$
This expression can be reproduced in the separable model of the
$\pi N$  $t$-matrix \cite{Ann,MF}

$$T({\bf p',p};E)
=\lambda g({\bf p'})g({\bf p})
(p_{\Delta}-p'_N)_{\alpha}
{{ u^{\alpha}({\bf p}_{\Delta})
{\overline u}^{\beta}({\bf p}_{\Delta})}\over{
{d}_{\Delta}(E) }}
(p_{\Delta}-p_N)_{\beta}\eqno(C.4c)$$

where
$${ d}_{\Delta}(E)=1-\lambda\int 
{{d^3{\bf q}}\over{(2\pi)^3}}
 {m_N\over{{2E_{\bf q}}_{\pi}{E_{\bf q}}_{N}}}
{{{\bf q}^2   g^2({\bf q}) }\over{E+io-{E_{\bf q}}_{\pi}-
{E_{\bf q}}_{N} }}\eqno(C.4d)$$
in the usual separable potential model and

$${d}_{\Delta}(E)=E-E_{\Delta}(bare)-
\Sigma_{\pi N}(E)\eqno(C.4e)$$
in the more complicated microscopic models with the bare energy 
$E_{\Delta}(bare)$ and the mass operator of the $\Delta$ resonance
 $\Sigma_{\pi N}(E)$.

Using the normalization condition \cite{Heller}, we get
$${d}_{\Delta}\Bigl(E=\sqrt{m_{\Delta}^2+{\bf P}_{\Delta}^2}\
\Bigr)=0.\eqno(C.4f)$$

Equations (C.4c) and (C.4d) can be represented in the form of  (C.4b)
with the corresponding redefinition of the form factors of 
the $\Delta$ resonances

$$|{ g}_{\Delta}(E)>\equiv
\biggl( {{ {d}_{\Delta}(E)}\over{E-E_{\Delta} }}\biggr)^{{1\over 2}}
{\sc U}(E) |\Psi_{\Delta}>
\eqno(C.4g)$$

In this way the expression  $E-E_{\Delta}(s)$ can be replaced by  
the propagator  ${d}_{\Delta}(E)$ which is constructed in 
the separable model.

$$E-E_{\Delta}(s)\Longrightarrow {d}_{\Delta}(E).\eqno(C.4h)$$

This allows  to separate the $\Delta$ 
pole and non-pole 
parts of the $\pi N$ amplitude in accordance with the  (A.9a,b,c,d).

The double $\Delta$ exchange term with the $\Delta-\gamma'\Delta'$ vertex
(3.21c) (Fig. 2B) can be extracted from the $\pi N$ bremsstrahlung amplitude 
$<out;{\bf p'}_{N}{\bf p'}_{\pi}|{\cal J}^{\mu}(0)|{\bf p}_{\pi}{\bf p}_N;in>$
(2.6) in the same way as in our previous papers \cite{Ann,MF,NP}.
Thus the  $s$-channel part
of the  full $\pi N$ bremsstrahlung amplitude
with the  double $\pi N$  intermediate states is

$$<out;{\bf p'}_{N}{\bf p'}_{\pi}|{\cal J}^{\mu}(0)|{\bf p}_{\pi}{\bf p}_N;in>
\Longrightarrow\sum_{ {\pi'' N''},{\pi'''N'''}}
\int d^4x' e^{ip'_{\pi}x'}d^4x e^{-ip_{\pi}x}
 <out;{\bf p'}_{N}|j_{\pi'}(x')\Bigr]|{\pi}'''N''';out>$$
$$\theta(x'_o)
<out;\pi'''N'''|{\cal J}^{\mu}(0)|{\pi'' N''};in>\theta(-x_o)
<in;{\pi'' N''}|j_{\pi}(x)|{\bf p}_N;in>\eqno(C.5a)$$

which after integration is transformed to

$$<out;{\bf p'}_{N}{\bf p'}_{\pi}|{\cal J}^{\mu}(0)|{\bf p}_{\pi}{\bf p}_N;in>
\Longrightarrow (2\pi)^6
\sum_{ {\pi'' N''},{\pi'''N'''}}
{{<out;{\bf p'}_{N}|{\sc U}(E_{\bf p'})|{ \Psi}_{\pi'' N''}({\bf p''})>}
\over{ {p'}^o_{N}+{p'}^o_{\pi}-{p'''}^o_{N}+{p'''}^o_{\pi}-i\epsilon}}$$
$$<out;\pi'''N'''|{\cal J}^{\mu}(0)|{\pi'' N''};in>
{{<{ \Psi}_{\pi'' N''}({\bf p''})|{\sc U}(E_{\bf p})|{\bf p}_N;in>}
\over{ {p}^o_{N}+{p}^o_{\pi}-{p''}^o_{N}+{p''}^o_{\pi}-i\epsilon}}
\eqno(C.5b)$$
where we used a connection between the $\pi N$ amplitude 
and the $\pi N$ wave function \cite{Ann,NP}
$$<out;{\bf p'}_{\pi}{\bf p'}_{N}|j_{\pi}(0)|{\bf p}_{N};in>=
<out;{\bf p'}_{N}|{\sc U}(E_{\bf p'})|{ \Psi}_{\pi N'}({\bf p})>,
\eqno(C.6)$$

Next we separate the $\pi N$ irreducible part 
$\biggl[<out;\pi'''N'''|{\cal J}^{\mu}(0)|{\pi'' N''};in>
\biggr]_{\pi\ N\ irreducible}$ of the full $\pi N$ bremsstrahlung amplitude
$<out;\pi'N'|{\cal J}^{\mu}(0)|{\pi N};in>$ as

$$<out;\pi'N'|{\cal J}^{\mu}(0)|{\pi N};in>=\sum_{ {\pi'' N''},{\pi'''N'''}}
<{ \Psi}_{\pi' N'}|(1-{\sc B})|\pi''' N''';out>$$
$$\biggl[<out;\pi'''N'''|{\cal J}^{\mu}(0)|{\pi'' N''};in>
\biggr]_{\pi\ N\ irreducible}<in;{\pi'' N''}|(1-{\sc B})|{ \Psi}_{\pi N}>,
\eqno(C.7)$$

Substituting (C.7) into (C.5b) and using (C.3) and (C.4a)
we get (3.6), where

$$
<{\bf P'}_{\Delta},{\sf m}_{\Delta}(s')|{\cal J}^{\mu}(0)
|{\bf P}_{\Delta},{\sf m}_{\Delta}(s)>=
\sum_{ {\pi'' N''},{\pi'''N'''}}
<{ \Psi}_{\Delta'}|(1-{\sc B})|\pi''' N''';out>$$
$$\biggl[<out;\pi'''N'''|{\cal J}^{\mu}(0)|{\pi'' N''};in>
\biggr]_{\pi\ N\ irreducible}<in;{\pi'' N''}|(1-{\sc B})|{ \Psi}_{\Delta}>,
\eqno(C.8)$$

and 

$$
<{\bf p'}_N,{\bf p'}_{\pi}|{\sl g}_{\pi' N'-\Delta'}|{\bf  P'}_{\Delta'}>=
<{\bf p'}_N,{\bf p'}_{\pi}|{\sc U}(E_{\bf p'})|{\Psi}_{\Delta'}({\bf P'})>,
\eqno(C.9a)$$

$$<{\bf P}_{\Delta}|{\sl g}_{\Delta-\pi N}|{\bf p}_N,{\bf p}_{\pi}>=
<{\Psi}_{\Delta}({\bf P})|{\sc U}(E_{\bf p})|{\bf p}_N,{\bf p}_{\pi}>
\eqno(C.9b)$$



\vspace{0.25cm}



\begin{thebibliography}{99}

\bibitem{Leung} K.C. Leung, M. Arman, H.C. Ballagh, Jr., P.F. Glodis, 
R.P. Haddock, B.M.K. Nefkens, and D.I. Sober, Phys. Rev. {\bf D14}
(1976) 698.

\bibitem{Nefkens} B. M. K. Nefkens at al., Phys. Rev. {\bf D18} (1978) 3911.

\bibitem{Wittman} R. Wittman, Phys. Rev. {\bf C37}
(1988) 2075.

\bibitem{Boss} A. M. Bosshard at al., Phys. Rev. {\bf D44} (1991) 1962;
 C. A. Meyer at al., Phys. Rev. {\bf D38} (1988) 754.





\bibitem{Low}  F. E. Low, Phys. Rev. {\bf 110}
(1958) 974.

\bibitem{Adler}  S. L. Adler, and Y. Dothan, Phys. Rev. {\bf 151}
(1966) 1267.

\bibitem{Kond}  L.A. Kondratyuk, and L.A. Ponomarev, Sov. Journal of 
Nucl. Phys.{\bf 7} (1968) 82.


\bibitem{Mink}  W. E. Fischer, and P. Minkowski, Nucl. Phys. {\bf B36}
(1972) 519.

\bibitem{Musa}  M. M. Musakhanov, Sov. J. Nucl. Phys. {\bf 19}
(1974) 319.



\bibitem{Pascual} P. Pascual, and R. Tarrach, Nucl. Phys. {\bf B134}
(1978) 133.


\bibitem{Liou1}  M. K. Liou, and Z. M. Ding, Phys. Rev. {\bf C35}
(1987) 651.




\bibitem{Ding} Z. M. Ding, D. Lin,  and M. K. Liou, Phys. Rev. {\bf C40}
(1989) 1291.

\bibitem{Liou2} D. Lin,  and M. K. Liou, Phys. Rev. {\bf C43}
(1991) R930.

\bibitem{Lin} D. Lin, M. K. Liou, and Z. M. Ding, Phys. Rev. {\bf C44}
(1991) 1819,


\bibitem{Beg} M. A.B. Beg, B.W. Lee, and A. Pais, Phys. Rev. Lett. {\bf 13}
(1964) 514,

\bibitem{Georgi} H. Georgi. Lie Algebras in
Particle Physics (Reading) 1982.

\bibitem{Pais} M. A.B. Beg, and A. Pais, Phys. Rev. {\bf 137}
(1965) B1514,

\bibitem{Brown} G. E. Brown, M. Rho, and V. Vento, Phys. Lett. {\bf B97}
(1980) 423.

\bibitem{Heller} L. Heller, S. Kumano, J. C. Martinez, and E. J. Moniz,
Phys. Rev. {\bf C35} (1987) 718.

\bibitem{Kriv}  M. I. Krivoruchenko, Sov. J. Nucl. Phys. {\bf 45}
(1987) 109.

\bibitem{Buch97}  A. J. Buchmann, E. Hern\'andez and Amand Faessler,
Phys. Rev. {\bf C55} (1997) 448.
\bibitem{Kim98} H.-C. Kim, M. Praszalowicz, and K. Goeke,
 Phys. Rev. {\bf D57} (1998) 2859.
\bibitem{Lin98} J. Linde, T. Ohlsson, and H. Snellman, 
 Phys. Rev. {\bf D57} (1998) 5916.
\bibitem{Acu98} A. Acus, E. Norvai${\check {\rm s}}$as, and D. O. Riska,
 Phys. Rev. {\bf C57} (1998) 2597.

\bibitem{Castro}  G. Lopez Castro, and I. A. Marino, 
Found. Phys. {\bf 23}(2003) 719; Nucl. Phys. {\bf 697} (2002) 440.

\bibitem{Franklin} J. Franklin, Phys. Rev. {\bf D66}
(2002) 033010.

\bibitem{BD1}  J. D. Bjorken and S.D.Drell, Relativistic Quantum Mechanics. 
(New York, Mc Graw-Hill) 1964.


\bibitem{BD2}  J. D. Bjorken and S.D.Drell, Relativistic Quantum Fields. (New
York, Mc Graw-Hill) 1965.


\bibitem{IZ}  C. Itzykson and C. Zuber. Quantum Field Theory. (New York,
McGraw-Hill) 1980.

\bibitem{NP}  A. I. Machavariani, Amand Faessler, and A. J. Buchmann.
 Nucl. Phys. {\bf A646} (1999) 231; (Erratum
{\bf A686} (2001) 601).
\bibitem{Ann} A. I. Machavariani, and Amand Faessler.
Ann. Phys. {\bf 309} (2004) 49.
\bibitem{MF} A. I. Machavariani, and Amand Faessler.
Phys. Rev.{\bf C72} (2005) 024002.


\bibitem{Amiri} M. El Amiri, G. Lopez Castro, and J. Pestieau. 
Nucl. Phys. {\bf A543} (1992) 673



\bibitem{PasVan2} Wen-Tai Chiang, M. Vanderhaeghen, Shin Nanan Yang
and D. Drechsel. Phys. Rev.{\bf C71} (2005) 15204


\bibitem{PasVan3} V. Pascalutsa, M. Vanderhaeghen, and Shin Nanan Yang.
Phys. Rept. {\bf 427} (2007) 125.



\bibitem{Brodsky1} S. Brodsky, and R.W. Brown. Phys. Lett.
{\bf 49} (1982) 966.


\bibitem{Brodsky2} S. Brodsky, and R.W. Brown. Phys. Rev.
{\bf D28} (1983) 624.








\bibitem{Nie} P. van Nieuwenhuizen, Phys. Rep.  {\bf 68} (1981) 189.
\bibitem{W} H.T.Williams, Phys. Rev. {\bf C31} (1985) 2297.
\bibitem{Bammer} M. Bemmerrouche, R. M.
Davidson, and N. C. Mukhopadhyay, Phys. Rev.  {\bf C39} (1989) 2339
 and references therein.

\bibitem{Fronsdal} R.E. Behrends, and C. Fronsdal, Phys. Rev.  
{\bf 106} (1957) 345.








\bibitem{Thomas} S. Theberge, A. W. Thomas and G. A. Miller, Phys. Rev.
{\bf D24} (1981) 216; {\bf D22} (1980) 2838.



\end{thebibliography}
\end{document}